\documentclass[12pt]{iopart}
\usepackage{iopams}
\usepackage{graphicx}
\expandafter\let\csname equation*\endcsname\relax
\expandafter\let\csname endequation*\endcsname\relax
\usepackage{amsmath}
\usepackage{amssymb}
\usepackage{cite}

\begin{document}

\title[Phase transition in   evolving  networks]
{Phase transition in   evolving   networks that combine preferential attachment and random node deletion}
\author{Barak Budnick, Ofer Biham and Eytan Katzav}
\address{
Racah Institute of Physics, 
The Hebrew University, 
Jerusalem 9190401, Israel
}
\eads{
\mailto{barak.budnick@mail.huji.ac.il},
\mailto{ofer.biham@mail.huji.ac.il}, 
\mailto{eytan.katzav@mail.huji.ac.il}
}

\begin{abstract}

Analytical results are presented for the structure
of networks that evolve via
a preferential-attachment-random-deletion (PARD) model
in the regime of overall network growth and in the regime of overall contraction. 
The phase transition between the two regimes is studied. 
At each time step a node addition and preferential attachment step takes place with
probability $P_{\rm add}$, and a random node deletion step 
takes place with probability 
$P_{\rm del} = 1 - P_{\rm add}$.
The balance between growth and contraction is 
captured by the parameter 
$\eta = P_{\rm add} - P_{\rm del}$,
which in the regime of overall network 
growth satisfies $0 < \eta \le 1$ and in the regime of overall network
contraction  
$-1 \le \eta < 0$.
Using the master equation and computer simulations
we show that for 
$-1 < \eta < 0$
the time-dependent degree distribution $P_t(k)$ converges towards
a stationary form $P_{\rm st}(k)$
which exhibits an exponential tail.
This is in contrast with the power-law tail of the stationary degree distribution
obtained for $0 < \eta \le 1$.
Thus, the PARD model has a phase transition at $\eta=0$,
which separates between two structurally distinct phases.
At the transition, for $\eta=0$, the degree distribution exhibits a stretched exponential tail. 
While the stationary degree distribution in the phase of overall growth
represents an asymptotic state, in the phase of overall contraction
$P_{\rm st}(k)$ represents an intermediate asymptotic state of a finite life span,
which disappears when the network vanishes.

\end{abstract}


\maketitle

\newpage

\section{Introduction}

Network theory provides a conceptual and mathematical framework for the analysis 
of structure and
dynamics in biological, social and technological networks
\cite{Dorogovtsev2003,Havlin2010,Newman2010,Estrada2011,Latora2017,Dorogovtsev2022}.
It was found that empirical networks are typically 
small-world networks whose diameters scale logarithmically with the network size.
Many of these networks exhibit
fat-tailed degree distributions with
scale free structures
\cite{Redner1998,Barabasi1999,Albert2002}. 
The growth processes that yield these structures were studied extensively.
It was found that the added nodes tend to connect preferentially 
to nodes of high degree, which are more visible or prominent in the network.
As a result, nodes of high degree tend to attract new connections at a higher rate.
Highly connected nodes (or hubs) often provide better access to information, resources or opportunities 
within the network. 
New nodes tend to benefit from these advantages by connecting to 
high degree nodes, further reinforcing their centrality.

The preferential attachment mechanism
gives rise to scale-free networks with power-law degree distributions
of the form
$P(k) \sim k^{- \gamma}$
\cite{Barabasi1999,Albert2002,Krapivsky2000,Dorogovtsev2000}.
These properties are captured by the
Barab\'asi-Albert (BA) model 
\cite{Barabasi1999}.
In this model, at each 
time step a new node is added to the network and
connects to $m$ of the existing nodes.
The probability 
of an existing node of degree $k$ to gain a link to the new node is proportional to $k$.
The degree distribution of the BA network exhibits a 
power-law tail
with $\gamma=3$.
Variants of the BA model were shown to yield power-law distributions with exponents in the range
$2 < \gamma \le 3$  
\cite{Krapivsky2000,Dorogovtsev2000,Bollobas2001}.
In scale-free networks of a finite size, there is always a maximum degree $k_{\rm max}$,
which is constrained by the network size $N$.
For $\gamma \le 2$ the mean degree $\langle K \rangle$ diverges in the limit of
$k_{\rm max} \rightarrow \infty$.
For $2 < \gamma < 3$ the mean degree is bounded while the second moment
$\langle K^2 \rangle$ diverges algebraically.
In the marginal case of $\gamma = 3$ the second moment diverges logarithmically,
while in the regime of $\gamma > 3$ it is bounded.

The robustness of complex networks is crucial for ensuring functionality, security and stability
of these networks.
The mathematical framework of percolation theory provides useful tools for the
analysis of network robustness against the deletion of nodes due to failures and attacks
\cite{Albert2000,Cohen2000,Cohen2001,Gao2015,Yuan2015,Shao2015,Havlin2015,Shekhtman2015,Shekhtman2016,Yuan2016,Muro2016,Vaknin2017}.
When the number of deleted nodes 
exceeds some threshold, the network disintegrates, resulting in a large number of disconnected components
\cite{Molloy1995,Molloy1998,Albert2000,Cohen2000,Cohen2001,Braunstein2016,Zdeborova2016}.
In this context, it was shown that scale-free networks are resilient to random failures and attacks targeting
random nodes 
\cite{Cohen2000}, 
but are vulnerable to 
targeted attacks aimed at high degree nodes or hubs 
\cite{Cohen2001}.

The evolution of the structure of networks that undergo contraction  
due to node deletion processes was recently studied
\cite{Tishby2019,Tishby2020}.
These processes include 
random, preferential  
and propagating node deletion.
The process of random node deletion describes random failures or random 
attacks that do not target any specific type of nodes.
The process of preferential node deletion describes attacks that
target high degree nodes, where the probability of a node to be selected for deletion
is proportional to its degree. 
Propagating node deletion
describes processes that propagate in the network from an infected node
to its neighbors.
Using the master equation for the degree distribution $P_t(k)$,
it was shown that for a broad class of network structures,
when the network contracts,  
its degree distribution evolves towards a Poisson distribution,
namely it becomes an Erd{\H o}s-R\'enyi (ER) network
\cite{Erdos1959,Erdos1960,Erdos1961}.
ER networks belong to an ensemble of maximum entropy 
random graphs
\cite{Coolen2017}.

The structure of networks that evolve under a combination 
of growth and contraction processes
was studied in Refs.
\cite{Moore2006,Bauke2011,Ghoshal2013},
which focus on 
the regime of
overall network growth.
More specifically, 
in the model studied by Moore et al.
\cite{Moore2006} the growth mechanism is based on the BA model
\cite{Barabasi1999},
where each new node 
is connected to $m$ existing nodes, which are selected preferentially,
while the contraction process takes place via random node deletion.
We thus refer to the model studied by Moore et al. 
\cite{Moore2006} 
as the Barab\'asi-Albert with random deletion (BARD) model.
It was found that in the regime of overall network growth the asymptotic degree distribution  
of the BARD model exhibits a power-law tail.
However, the complementary regime, in which the overall process is of network contraction
has not been studied. 

The case of overall network contraction occurs when the rate of node deletion exceeds the
rate of node addition. For example, social networks contract when the number of users that
leave the network is larger than the number of users that join
\cite{Torok2017,Lorincz2019}.
A similar situation may occur in networks of business firms
in which the nodes represent individual companies, while the edges describe the
connections between them.
These networks evolve in time as companies are launched, grow, mature and 
may eventually decline and dissolve.
A remarkable study of a declining network was done by Saavedra et al.
\cite{Saavedra2008},
who analyzed the business network of the
New York City garment industry, in which 
the nodes represent designers and contractors and
the links represent coproductions of lines of clothing.
The analysis is focused on the period of 19 years between 1985 and 2003,
during which the network size had shrunk more than 10-fold.
In spite of the dramatic decline in size, the network remained topologically 
robust in the sense that it resisted quick fragmentation and
the degree distribution remained stationary, maintaining its truncated
power-law form. This lasted until around the year 2000, when finite size effects became more
pronounced. This robustness enabled the network to maintain its functionality in spite of
the declining size. Using computer modeling and simulations, Saavedra et al.
concluded that the deletion process is anti-preferential, namely the
probability of a node to be deleted is inversely proportional to its degree.
In simple terms it implies that the least connected nodes are the most vulnerable to
deletion from the network.
They also found that the deletion process is accompanied by a 
partial recovery process via preferential attachment.
The robustness of the degree distribution as the network declines provides
a strong motivation for the study of stationary solutions for the degree 
distribution of contracting networks.
 
In this paper we analyze the emerging structure of networks that evolve
under a combination of growth, by node addition and preferential attachment,
and contraction, by random node deletion,
in the regime of overall network growth and in the regime of overall contraction.
The phase transition between the two regimes is studied.
Unlike the BARD model, in the model studied here, the new nodes added to the network are initially isolated.
They gain links in subsequent time steps via the preferential attachment process,
that connects pairs of already existing nodes in the network.
This model is referred to as the preferential-attachment-random-deletion (PARD) model.

Using the master equation  
we show that in the regime of overall network contraction,
the degree distribution converges towards
a stationary form $P_{\rm st}(k)$, which exhibits an exponential tail.
This is in contrast with the case of growing networks for which
the distribution $P_{\rm st}(k)$ is a
power-law distribution.
Thus, the PARD model has a phase transition between the growing phase and the contracting phase,  
where the stationary degree distribution changes in a qualitative way.
At the transition, the network maintains its original size, apart from fluctuations, and
the stationary degree distribution $P_{\rm st}(k)$ follows
a stretched exponential form.
It is found that the stationary degree distribution $P_{\rm st}(k)$ of the PARD model
is smooth and follows the same functional form over the whole range of degrees $k \ge 0$.
This is in contrast with the BARD model that exhibits a cusp at $k=m$.
This cusp separates between the regime of high degree nodes and the regime of low degree nodes.
The steady state degree distribution $P_{\rm st}(k)$ behaves differently in the two regimes.
These features make the PARD model somewhat easier to analyze than the BARD model.

The results obtained for the PARD model are
in contrast with the random-attachment-random-deletion (RARD) model, 
in which $P_{\rm st}(k)$ the same Poisson-like tail both in the overall growth and in the overall contraction phases
with no phase transition between them
\cite{Budnick2022}.
Another model, which is similar to the RARD model in the case in which the network 
size remains fixed, was recently studied 
\cite{Artime2022}.
It was also found to exhibit a Poisson-like tail.

The paper is organized as follows. 
In Sec. 2 we present the PARD model.
In Sec. 3 we present the master equation for the
degree distribution $P_t(k)$ of the PARD model and
obtain a differential equation for the generating function 
$G_t(u)$.
In Sec. 4 we calculate the stationary degree distribution
$P_{\rm st}(k)$ and show that there is a phase transition between the
regimes of overall growth and overall contraction.
In Sec. 5 we briefly consider the corresponding transition in the BARD
model.
The results are discussed in Sec. 6 and summarized in Sec. 7.
In Appendix A 
we calculate the derivatives of the stationary generating function $G(u)$ 
in the special case of $\eta=0$.
In Appendix B we calculate the stationary degree distribution $P_{\rm st}(k)$
of the BARD model.

\section{The model}

Starting from an initial network that consists of $N_0$ nodes at time $t=0$,
in the PARD model the network evolves according to the following procedure.
At each time step starting from $t=1$, one of two possible processes takes place:
(a) {\it growth step}: with probability $P_{\rm add}$, an isolated node (of degree $k=0$)
is added to the network. This is followed by the addition of $m$ edges,
by repeating the following step $m$ times:
each time a node, which is selected uniformly at random, is connected 
to another node, which is selected preferentially 
(namely, with a probability that is proportional to its degree),
under the condition that the two nodes have not been connected  before;
(b) {\it contraction step}: with probability
$P_{\rm del}=1-P_{\rm add}$,
a random node is deleted, together with its edges.

In case that a growth step is selected at time $t$, the size of the network increases
to $N_{t+1}=N_t + 1$.
In each one of the subsequent edge addition steps,
the degrees of both the randomly selected node
and of the preferentially selected node are increased by $1$.
When a contraction step is selected at time $t$, the size of the network  
decreases to $N_{t+1}=N_t - 1$.
Consider a node of degree $k$, whose neighbors are of
degrees $k'_r$, $r=1,2,\dots,k$.
When such node is deleted,
the degrees of its neighbors are reduced to
$k'_r-1$, $r=1,2,\dots,k$. 

The expected number of nodes in the network at time $t$ is given by
 
\begin{equation}
N_t = N_0 + \eta t,
\label{eq:Nt}
\end{equation}

\noindent
where the growth/contraction rate
 
\begin{equation}
\eta = P_{\rm add} - P_{\rm del} 
\end{equation}

\noindent
represents the expectation value of the change in the size of the network at each time step.
The parameter $\eta$ provides a useful classification of the possible scenarios
of growth and contraction.
The case of pure growth is described by $\eta=1$.
For $0 <\eta < 1$ the overall process is of network growth, 
while for $-1 \le \eta <0$ the overall
process is of network contraction.
In the special case of $\eta=0$ the network size remains the same, apart from 
stochastic fluctuations.
In the analysis below, it will be useful to express the probabilities $P_{\rm add}$ and $P_{\rm del}$ 
in terms of the parameter $\eta$, namely

\begin{equation}
P_{\rm add} =  \frac{1+\eta}{2} 
\label{eq:Padd}
\end{equation}

\noindent
and

\begin{equation}
P_{\rm del} = \frac{1-\eta}{2}.
\label{eq:Pdel}
\end{equation}

\noindent
In the case of 
overall network contraction, namely for
$-1 < \eta < 0$, it is useful to define the normalized time variable

\begin{equation}
\tau = \frac{ | \eta | t }{N_0},
\label{eq:tau}
\end{equation}

\noindent
which measures the expected fraction of nodes that are deleted from the network
up to time $t$. The expected size of the contracting network at time $t$ 
can be expressed by $N_t = N_0(1-\tau)$.
Note that the network vanishes at $\tau=1$, which occurs at time

\begin{equation}
t_{\rm vanish} = \frac{N_0}{|\eta|}.
\label{eq:t_vanish}
\end{equation}

In the PARD model, the $m$ edges added at each growth step
connect pairs of existing nodes, such that one
end-node of each edge is selected uniformly at random and the other end-node
is selected preferentially, namely with probability that is proportional to the
degree of the node. 
The degrees of the random end-nodes may take values in the range $k \ge 0$,
while the degrees of the preferential end-nodes may take values in the range $k \ge 1$. 
An efficient way to implement the preferential selection of such end-node
is by selecting a random edge and then choosing randomly one of the
two end-nodes of the selected edge.

The PARD model is motivated by the dynamics of online social networks.
In these networks the formation of a friendship between two users involves a
two-step process, in which one user initiates the connection (by sending a friendship request)
and the other user confirms it (by accepting the request).
In spite of this asymmetric process, the resulting connection is mutual and symmetric,
and is thus represented by an undirected edge.
The user who initiated the connection is represented by a node selected uniformly at random,
while the user who accepted the friendship request is represented by a node selected preferentially.
This is due to the fact that users who have many friends are more likely to receive additional friendship
requests than users who have few friends.
The model also reflects the fact that users in online social networks typically initiate new friendships
gradually over time rather than immediately upon joining the network.
This process reflects real-world social dynamics, where relationships build and evolve over time.
The contraction step in the PARD model accounts for users who deactivate or delete
their social media accounts. This may be due to loss of
interest, concerns about privacy or due to their migration to other social networks
\cite{Torok2017,Lorincz2019}.

In the BA model
the degree of each new node upon its addition to the
network is $k=m$, which later increases as it gains more links. 
This model is motivated by the growth dynamics of the World Wide Web,
in which the initial $m$ links of a new webpage represent outgoing hyperlinks from the new webpage
to existing webpages.
The connections that are gained at later times represent incoming hyperlinks from
younger webpages. For simplicity, these hyperlinks are represented by undirected edges.
Due to these growth rules, the degrees of all the nodes in the BA model satisfy $k \ge m$.
As a result, the degree distribution of the BARD model 
is divided into two regimes:
the regime of high degree nodes ($k > m$), 
which gained more links than they lost, and
the regime of low degree nodes ($k < m$), 
which lost more links than they gained.
The two regimes are separated by a cusp at $k=m$
\cite{Moore2006}.
In the PARD model the new nodes are added as an isolated nodes
of degree $k=0$ and gain links in a stochastic manner in subsequent time steps.
Since these links are added one at a time, the resulting degree distribution 
exhibits the same functional form over the entire range of
possible values of the degree $k$.

\section{The master equation}

Consider an ensemble of 
networks of size $N_0$ at time $t=0$,
whose initial degree distribution is given by $P_0(k)$.
While the formalism applies to any form of the initial degree distribution,
it is often convenient to start from an ER network or a BA network.
The networks evolve under a combination of growth,
via node addition and preferential attachment,
and contraction, via random node deletion.
Below we derive a master equation 
\cite{Vankampen2007,Gardiner2004}
that describes the time
evolution of the degree distribution 

\begin{equation}
P_t(k) = \frac{N_t(k)}{N_t},
\label{eq:P_t(k)}
\end{equation}

\noindent
where $N_t(k)$, $k=0,1,\dots$, is the number of nodes of degree $k$
in the network at time $t$
and $N_t = \sum_k N_t(k)$ is the network size at time $t$.

The master equation accounts for the time 
evolution of the degree distribution $P_t(k)$ over an ensemble of networks.
All the network instances in the ensemble have
the same initial size $N_0$ and their degree sequences are drawn from the same initial degree distribution $P_0(k)$.
To derive the master equation, we first consider the
time evolution of $N_t(k)$, which is expressed in terms
of the forward difference

\begin{equation}
\Delta_t N_t(k) = N_{t+1}(k) - N_t(k).
\end{equation}

\noindent
Following the derivation presented in Ref.
\cite{Budnick2022},
we obtain a difference equation of the form

\begin{eqnarray}
\Delta_t N_t(k) &=&
P_{\rm add} \delta_{k,0}
+
m P_{\rm add}  \frac{  N_t(k-1) - N_t(k)  }{N_t} 
\nonumber \\
&+&  
m P_{\rm add}    \frac{ (k-1) N_t(k-1) - k N_t(k)  }{\langle K \rangle_t N_t}
\nonumber \\
&+&
P_{\rm del}
\frac{(k+1)[ N_t(k+1) -  N_t(k) ]}{N_t},
\label{eq:DeltaNtkde0}
\end{eqnarray}

\noindent
where $\delta_{i,j}$ is the Kronecker delta.
The first term on the right hand side of Eq. (\ref{eq:DeltaNtkde0})
accounts for the addition of an isolated node.
The second term accounts for the increased degree
of the end-node selected randomly, while the third term accounts for the increased degree
of the end-node selected preferentially.
The last term accounts for the process of random node deletion.
We use the convention in which the boundary condition $N_t(-1)=0$ 
is imposed for all values of $t$.

Since nodes are discrete entities the processes of node addition and 
deletion are intrinsically discrete. Therefore, the replacement of the
forward difference $\Delta_t N_t(k)$ 
on the left hand side of Eq. (\ref{eq:DeltaNtkde0})
by a time derivative of the form
$dN_t(k)/dt$ involves an approximation.
The error associated with this approximation was shown to be
of order $1/N_t^2$, which quickly vanishes for sufficiently large networks
\cite{Tishby2019}.
Therefore, the difference equation (\ref{eq:DeltaNtkde0}) can be replaced by a
differential equation.

To complete the derivation of the master equation we take the
time derivative of Eq. (\ref{eq:P_t(k)}), which is given by

\begin{equation}
\frac{d P_t(k) }{dt}  = 
\frac{1}{N_t} \frac{d N_t(k)}{dt}  - \frac{N_t(k)}{N_t^2} \frac{d N_t}{dt} .
\label{eq:dPt_Nt}
\end{equation}

\noindent
Inserting the time derivative $d N_t(k)/dt$ from Eq. (\ref{eq:DeltaNtkde0}) 
and using the fact that
$d N_t/dt=\eta$ 
[from Eq. (\ref{eq:Nt})],
we obtain the master equation 
\cite{Vankampen2007,Gardiner2004}

\begin{eqnarray}
\frac{d P_t(k)}{dt}  &=&
\frac{1+\eta}{2 N_t} [ \delta_{k,0} - P_t(k) ]
+ \frac{ m (1+\eta) }{2 N_t} [P_t(k-1)-P_t(k)]
\nonumber \\
&+&
\frac{ m (1+\eta) }{2 \langle K \rangle_t N_t} [(k-1) P_t(k-1) - k P_t(k)]
\nonumber \\
&+& \frac{1-\eta}{2 N_t}
\left[ (k+1)P_t(k+1) - k P_t(k) \right],
\label{eq:dP(t)/dtRC0}
\end{eqnarray}

\noindent
where we have also expressed $P_{\rm add}$ and $P_{\rm del}$
in terms of $\eta$, using Eqs. (\ref{eq:Padd}) and (\ref{eq:Pdel}).
The master equation consists of a set of coupled ordinary differential equations
for $P_t(k)$, $k=0,1,2,\dots$.
Note that the second term in the first square bracket on the right hand side 
of Eq. (\ref{eq:dP(t)/dtRC0}) is a result of the transformation from $N_t(k)$
to $P_t(k)$. 
More specifically, it comes from the second term on the right hand side of 
Eq. (\ref{eq:dPt_Nt}).

We denote the time-dependent generating function by

\begin{equation}
G_t(u) = \sum_{k=0}^{\infty} u^k P_t(k),
\end{equation}

\noindent
which is the Z-transform of the time-dependent degree distribution $P_t(k)$
\cite{Phillips2015}.
Multiplying Eq. (\ref{eq:dP(t)/dtRC0}) by $u^k$ and summing up over $k$,
we obtain a partial differential equation for the generating function,
which is given by

\begin{eqnarray}
N_t \frac{ \partial G_t(u) }{\partial t}  &=&
\frac{1+\eta}{2} \left[ 1 - G_t(u) \right]
+  \frac{m(1+\eta)}{2} (u-1) G_t(u)
\nonumber \\
&+& \frac{m(1+\eta)}{2 \langle K \rangle_t} (u-1) \frac{ \partial G_t(u) }{\partial u}
+ \frac{1 - \eta}{2} (1-u) \frac{ \partial G_t(u) }{\partial u}.
\label{eq:diffeq0z}
\end{eqnarray}

\noindent
Rearranging terms in Eq. (\ref{eq:diffeq0z}), we obtain

\begin{eqnarray}
(N_0 + \eta t) \frac{ \partial G_t(u) }{\partial t} 
&+& \left[    \frac{m (1+\eta)   }{2 \langle K \rangle_t} u - \frac{1-\eta}{2}  \right]
(1-u)
\frac{ \partial G_t(u) }{\partial u}
\nonumber \\
&+& 
\frac{1+\eta}{2} \left[ m(1-u) + 1 \right] G_t(u) 
= \frac{1+\eta}{2}.
\label{eq:diffeq0a}
\end{eqnarray}

\noindent
This is a first order inhomogeneous partial differential equation of two variables.
Note that the time dependent mean degree $\langle K \rangle_t$ 
appears as a coefficient in Eq. (\ref{eq:diffeq0a}).
Since the mean degree depends on the time dependent degree distribution
$P_t(k)$ itself, the appearance of $\langle K \rangle_t$ in Eq. (\ref{eq:diffeq0a})
makes the equation nonlinear. 

The mean degree of the network at time $t$ is given by

\begin{equation}
\langle K \rangle_t = \frac{ d G_t(u) }{d u} \bigg\vert_{u=1}.
\end{equation}

\noindent
Differentiating equation (\ref{eq:diffeq0a}) with respect to $u$ and
inserting $u=1$, we obtain

\begin{equation}
(N_0 + \eta t) \frac{ d \langle K \rangle_t }{dt}   = 
- \langle K \rangle_t + m(1+\eta).
\label{eq:Ktmean}
\end{equation}

\noindent
Solving this equation, we obtain an explicit expression for the time dependent mean degree:

\begin{equation}
\langle K \rangle_t = 
\left\{
\begin{array}{ll}
  \left( 1 +  \frac{\eta t}{N_0} \right)^{ - \frac{ 1 }{ \eta} }  
  \langle K \rangle_0
  + \left[ 1 - \left( 1 +  \frac{\eta t}{N_0} \right)^{ - \frac{ 1 }{ \eta} } \right]
  m(1+\eta)   & 0 < \eta \le 1 \\
e^{ - \frac{ t }{ N_0} } \langle K \rangle_0 + 
\left(1 - e^{ - \frac{ t }{ N_0} } \right) m
&  \eta = 0 \\ 
   \left( 1 -  \frac{|\eta| t}{N_0} \right)^{   \frac{ 1 }{ |\eta|} }  
  \langle K \rangle_0
  + \left[ 1 - \left( 1 - \frac{|\eta| t}{N_0} \right)^{  \frac{ 1 }{ |\eta|} } \right]
  m(1 - |\eta|)
      & -1 < \eta < 0,
\end{array}
\right.  
\label{eq:K_t}
\end{equation}

\noindent
where the first two lines on the right hand side of Eq. (\ref{eq:K_t}) hold for $t \ge 0$,
while the third line holds for $0 < t < t_{\rm vanish}$.
Note that Eq. (\ref{eq:K_t}) is identical to the result for $\langle K \rangle_t$
in the RARD model (Eq. (44) in Ref. \cite{Budnick2022}).
For $0 \le \eta \le 1$, in the long time limit, the mean degree converges towards its 
stationary value, given by

\begin{equation}
\langle K \rangle_{\rm st} = m(1+\eta),
\label{eq:Kst}
\end{equation}

\noindent
as the network
continues to grow for an unlimited period of time.
In contrast, in the regime of overall network contraction, 
the mean degree $\langle K \rangle_t$ converges towards
$\langle K \rangle_{\rm st}$
while the network contracts and eventually vanishes.
In all the cases of Eq. (\ref{eq:K_t}), the time scale for the convergence
of $\langle K \rangle_t$ towards $\langle K \rangle_{\rm st}$
is $| \eta | t_{\rm vanish} = N_0$.
In the case of overall network contraction, this implies that
the stationary state exists in the time window between
$| \eta | t_{\rm vanish}$ and $t_{\rm vanish}$.
This time window becomes wider as $| \eta |$ is decreased.
Note the remarkable result that $\langle K \rangle_{\rm st}$ does
not depend on the specific form of $P_t(k)$ and only on the initial mean degree $\langle K \rangle_0$.

\section{The stationary degree distribution $P_{\rm st}(k)$}

In the long time limit the generating function $G_t(u)$ 
converges towards an asymptotic form that satisfies
$\partial G_t(u)/\partial t = 0$,
which can be expressed by

\begin{equation}
G(u) = \sum_{k=0}^{\infty} u^k P_{\rm st}(k).
\label{eq:Gu}
\end{equation}

\noindent
In this limit the mean degree $\langle K \rangle_t$
converges towards
$\langle K \rangle_{\rm st}$.
In this case, the differential equation (\ref{eq:diffeq0a}) for the generating function
can be simplified using Eq. (\ref{eq:Kst}),
and is reduced to

\begin{equation}
(\eta + u - 1)
(1-u)
\frac{ d G(u) }{d u}
+ (1+\eta)  \left[ m(1-u) + 1 \right] G(u) 
=  1+\eta .
\label{eq:diffeq0b}
\end{equation}

\noindent
This equation exhibits two singular points, one at 
$u=1$ and the other at $u=1-\eta$.
Note the plugging $u=1$ into Eq. (\ref{eq:diffeq0b}), recovers the normalization condition
for $P_{\rm st}(k)$ and thus does not add new information.
As long as $\eta \ne 0$ 
the points $u=1$ and $u=1-\eta$ 
are two distinct regular-singular points
\cite{Bender1999}.
Therefore, the radius of convergence of a series expansion for $G(u)$ around $u=0$
is determined by the value of $\eta$.
More specifically, the radius of convergence $R$ is equal to the 
distance from the origin of the singularity that is nearest to the origin.
Thus, for $0 < \eta < 1$ the radius of convergence is $R=1-\eta$,
while for $-1 < \eta < 0$ it is $R=1$.
In the special case of $\eta=0$ the two singularities coalesce and become
a single irregular-singular point at $u=1$.
These considerations suggest that the irregular-singular point at $\eta=0$
represents a phase transition that
separates between the domains of positive and negative values of $\eta$,
which are of a different nature.
 
\subsection{The case of $0 < \eta \le 1$}
 
Here we solve Eq. (\ref{eq:diffeq0b}) in the case of overall network growth,
namely for $0 < \eta \le 1$.
To this end we express the stationary generating function
in the form

\begin{equation}
G(u) = G^{(p)}(u) + G^{(h)}(u),
\end{equation}

\noindent
where $G^{(h)}(u)$ is the solution of the homogeneous equation
and $G^{(p)}(u)$ is a solution of the inhomogeneous equation.
The homogeneous equation is given by

\begin{equation}
\frac{ d G^{(h)}(u) }{du} =\frac{1+\eta}{1-\eta-u} 
\left( m + \frac{1}{1-u} \right) G^{(h)}(u).
\end{equation}

\noindent
The homogeneous solution is

\begin{equation}
G^{(h)}(u) = C (1-u)^{ 1 + \frac{1}{\eta} }
(1-\eta-u)^{ -  m(1+\eta) - 1 - \frac{1}{\eta} },
\label{eq:Ghu0eta1}
\end{equation}

\noindent
where $C$ is an integration constant.
The homogeneous solution diverges at $u=1-\eta < 1$,
which is inconsistent with the fact that $G(u)$ is a probability 
generating function that should be bounded in the range
$0 < G(u) < 1$ for 
$0 < u < 1$.
This implies that $C=0$. Thus, the solution for $G(u)$ consists
only of the inhomogeneous term $G^{(p)}(u)$.

We now turn to the solution of the inhomogeneous equation.
Multiplying Eq. (\ref{eq:diffeq0b}) by the integration factor

\begin{equation}
M(u) = (1-\eta-u)^{m(1+\eta)+1+\frac{1}{\eta}} (1-u)^{ -  1 - \frac{1}{\eta} },
\end{equation}

\noindent
one obtains

\begin{equation}
\frac{d}{du} [ G(u) M(u) ] = - (1+\eta) (1-\eta-u)^{ m(1+\eta) + \frac{1}{\eta} }
(1-u)^{ -  2 - \frac{1}{\eta}   }.
\label{eq:duGM}
\end{equation}

\noindent
Integrating Eq. (\ref{eq:duGM}) between $u$ and $1-\eta$
and using the fact that $M(1-\eta)=0$, we obtain

\begin{equation}
G(u) M(u) = (1+\eta) \int_{u}^{1-\eta}
(1-\eta-v)^{ m(1+\eta) + \frac{1}{\eta} }
(1-v)^{ - 2 - \frac{1}{\eta} } dv.
\end{equation}

\noindent
Carrying out the integration, we find that

\begin{equation}
G(u) = \frac{1}{m \eta + 1} \left( \frac{1-u}{\eta} \right)^{ 1 + \frac{1}{\eta} }
\, _2F_1 \left[ \left.
\begin{array}{c}
2+ \frac{1}{\eta}, m(1+\eta)+1+\frac{1}{\eta} \\
m(1+\eta)+2+\frac{1}{\eta}
\end{array}
\right|    1 - \frac{1-u}{\eta}
\right],
\label{eq:Gpu0eta1}
\end{equation}

\noindent
where 

\begin{equation}
_2F_1 \left[ \left.
\begin{array}{c}
a, b \\
c
\end{array}
\right| z 
\right] =
\sum_{n=0}^{\infty} 
\frac{ (a)_n (b)_n }{ (c)_n } \frac{ z^n }{ n! } 
\label{eq:2F1}
\end{equation}

\noindent
is the hypergeometric function 
and

\begin{equation}
(x)_n = 
\left\{
\begin{array}{ll}
1 & n=0 \\
\prod\limits_{i=0}^{n-1} (x+i) & n \ge 1
\end{array}
\right.
\end{equation}

\noindent
is the Pochhammer symbol
\cite{Olver2010}.
Note that the hypergeometric function  
$_2F_1[a,b;c;z]$ 
is undefined if $c$ is a non-positive integer. 
We have verified that in all the expressions that include the hypergeometric function, 
$c$ does not take any non-positive integer values, 
ensuring that the hypergeometric function is well-defined.

The stationary degree distribution is 
obtained by differentiating the generating function $G(u)$

\begin{equation}
P_{\rm st}(k) = \frac{1}{k!} \frac{ d^k G(u) }{ d u^k }  \bigg\vert_{u=0}.
\label{eq:PtkD}
\end{equation}
 
\noindent
Using identity 15.5.6 from Ref. \cite{Olver2010},
we obtain

\begin{eqnarray}
P_{\rm st}(k) &=& \frac{1}{m \eta + 1} \left( \frac{1}{\eta} \right)^{1+\frac{1}{\eta}}
\frac{ (m + m \eta)_k }{ \left( m + m \eta + 2 + \frac{1}{\eta} \right)_k }
\times
\nonumber \\
& &
\, _2F_1 \left[ \left.
\begin{array}{c}
2+ \frac{1}{\eta}, m(1+\eta)+1+\frac{1}{\eta} \\
m(1+\eta)+2+\frac{1}{\eta} + k
\end{array}
\right|    1 - \frac{1}{\eta}
\right].
\label{eq:Pketapos}
\end{eqnarray}

\noindent
In the large $k$ limit, 
the hypergeometric function on the right hand side of Eq. (\ref{eq:Pketapos})
is dominated by the $n=0$ term in the sum on the right hand side of Eq. (\ref{eq:2F1}),
which is equal to $1$.
Approximating the ratio between the two Pochhammer symbols on the right hand side of 
Eq. (\ref{eq:Pketapos}) using the Stirling formula, we obtain 

\begin{equation}
P_{\rm st}(k) \simeq \frac{1}{m \eta + 1} \left( \frac{1}{\eta} \right)^{1+\frac{1}{\eta}}
\frac{ \Gamma \left( m + m \eta + 2 + \frac{1}{\eta} \right)  }{ \Gamma(m + m \eta)  }
k^{ -   2 - \frac{1}{\eta} },
\label{eq:Pketapostail}
\end{equation}

\noindent
where $\Gamma(x)$ is the Gamma function
\cite{Olver2010}.
This result is consistent with the asymptotic results obtained in Refs. 
\cite{Moore2006,Ghoshal2013,Bauke2011}.

In the special case of pure network growth, where $\eta=1$,
the degree distribution is reduced to

\begin{equation}
P_{\rm st}(k) = \frac{ 4 m (2m+1) }{(k+2m+2)(k+2m+1)(k+2m)}.
\label{eq:PkBA}
\end{equation}

\noindent
This result resembles the degree distribution 
of the Barab\'asi-Albert model of network growth
with preferential attachment
\cite{Krapivsky2000,Dorogovtsev2000}.
Note that the second moment of the degree distribution for $\eta=1$, given by
Eq. (\ref{eq:PkBA}),
diverges logarithmically.
This is in contrast to the case of $0 < \eta < 1$,
in which the second moment of the degree distribution,
given by Eq. (\ref{eq:Pketapos}), is finite.
This may be attributed to the fact that at $\eta=1$
the regular-singular point of the differential equation 
(\ref{eq:diffeq0b}) at $u=1-\eta$ converges towards $u=0$.
Since the generating function $G(u)$ is defined as a power series around
$u=0$ the existence of such a Taylor expansion is not guaranteed.
Our solution shows that such an expansion indeed exists, but its
analytical properties are not the same as obtained for $\eta<1$.

\begin{figure}
\centerline{
\includegraphics[width=7.0cm]{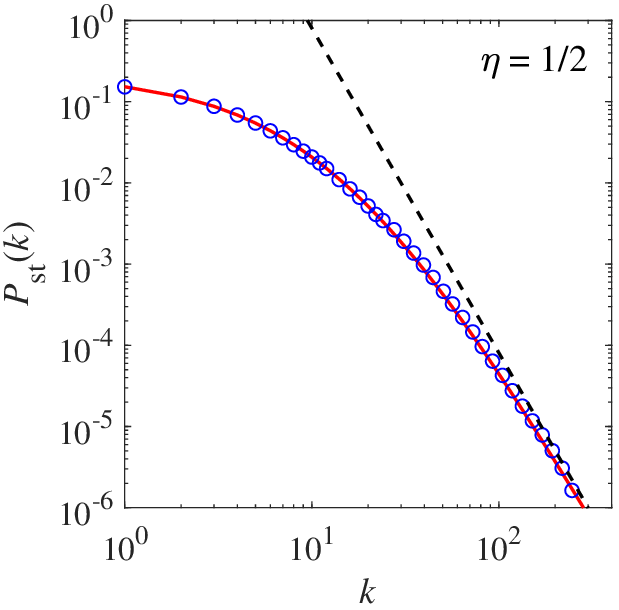} 
}
\caption{
Analytical results (solid line), 
obtained from Eq. (\ref{eq:Pketapos}),
for the stationary
degree distributions $P_{\rm st}(k)$ of a network that evolves via the PARD model
in the regime of overall network growth with $\eta=1/2$
and $m=4$.
We also present simulation results (circles) for the time-dependent degree distribution
$P_t(k)$ of such network that has grown from an initial
ER network of size $N_0=100$ with mean degree of $c=3$, up to a size of $N=10^4$.
The simulation results for $P_t(k)$ coincide with the analytical results for $P_{\rm st}(k)$,
which implies that the degree distribution has fully converged to its asymptotic form.
In the large $k$ limit, the stationary degree distribution $P_{\rm st}(k)$ converges towards
the asymptotic power-law tail (dashed line), given by Eq. (\ref{eq:Pketapostail}).
}
\label{fig:1}
\end{figure}

In Fig. \ref{fig:1} we present analytical
results (solid lines), 
obtained from Eq. (\ref{eq:Pketapos}),
for the stationary
degree distributions $P_{\rm st}(k)$ of networks that evolve 
via the PARD dynamics
for 
$\eta=1/2$.
We also present simulation results (circles) for the 
time-dependent degree distribution $P_t(k)$ at $N=10^4$.
The initial network used in the simulations 
is an ER network of size $N_0=100$ with mean degree $c=3$.
The simulation results for $P_t(k)$ are in very good agreement 
with the analytical results for $P_{\rm st}(k)$.
The dashed line shows the asymptotic power-law tails, given by 
Eq. (\ref{eq:Pketapostail}).

\subsection{The case of $\eta=0$}

Specializing to the case of
$\eta=0$ in Eq. (\ref{eq:diffeq0b}), we obtain

\begin{equation}
  -(1-u)^2
\frac{ d G(u) }{d u}
+   \left[ m(1-u) + 1 \right] G(u) 
=  1.
\label{eq:diffeq0d}
\end{equation}

\noindent
Interestingly, the two singularities of Eq. (\ref{eq:diffeq0b}) coincide
in the case of $\eta=0$ and become a quadratic singularity at $u=1$.
This is an indication for the existence of a phase transition at $\eta=0$.
The solution of the homogeneous equation is given by

\begin{equation}
G^{(h)}(u) = C (1-u)^{-m} \exp \left( \frac{1}{1-u} \right),
\label{eq:Ghu0eta0}
\end{equation}

\noindent
where $C$ is an integration constant.
This solution diverges at $u=1$. Since $G(1)$ should be finite, as it is a probability generating function, this solution is discarded.

Multiplying Eq. (\ref{eq:diffeq0d}) by the integration factor

\begin{equation}
M(u) =   (1-u)^{m} \exp \left( - \frac{1}{1-u} \right),
\end{equation}

\noindent
and integrating both sides between $u$ and $1$, we obtain

\begin{equation}
G(u) = (1-u)^{-m} \exp \left( \frac{1}{1-u} \right) 
\int_{u}^{1} (1-v)^{m-2} \exp \left( - \frac{1}{1-v} \right) dv.
\end{equation}

\noindent
This generating function can be expressed in the form

\begin{equation}
G(u) = (1-u)^{-m} \exp \left( \frac{1}{1-u} \right) 
\Gamma \left( -m+1,\frac{1}{1-u} \right),
\label{eq:Gueta0}
\end{equation}

\noindent
where $\Gamma(a,z)$ is the upper incomplete gamma function
\cite{Olver2010}.

In the analysis below we insert $G(u)$ from 
Eq. (\ref{eq:Gueta0}) into Eq. (\ref{eq:PtkD}) and extract the degree distribution $P_{\rm st}(k)$.
To this end, in Appendix A we show that the $k$'th derivative of $G(u)$
can be expressed in the form

\begin{equation}
\frac{ d^k G(u) }{du^k}  =
(m)_k \exp \left( \frac{1}{1-u} \right)
\sum_{i=0}^{k} (-1)^i \binom{k}{i} 
\Gamma \left( -m+1-i, \frac{1}{1-u} \right)
\left( \frac{1}{1-u} \right)^{m+k+i},
\label{eq:dkGu}
\end{equation}

\noindent
where $\binom{k}{i}$ is the binomial coefficient.
Inserting $d^k G(u)/du^k$ from Eq. (\ref{eq:dkGu})
into Eq. (\ref{eq:PtkD}), we obtain

\begin{equation}
P_{\rm st}(k) = 
\frac{e}{k!}
(m)_k  
\sum_{i=0}^{k} (-1)^i \binom{k}{i} 
\Gamma \left( -m+1-i, 1 \right).
\label{eq:Pketa00}
\end{equation}

\noindent
Replacing the incomplete Gamma function by 
its integral representation, exchanging the order
of the sum and the integral and carrying out the 
summation, we obtain

\begin{equation}
P_{\rm st}(k) = 
\frac{e}{k!}
(m)_k  
\int_{1}^{\infty} t^{-m} e^{-t} \left( 1 - \frac{1}{t} \right)^k dt.
\label{eq:Pketa0b}
\end{equation}

\noindent
Replacing the integration variable by $z=t-1$ and using identity
13.4.4 in Ref. \cite{Olver2010}, we obtain

\begin{equation}
P_{\rm st}(k) =  (m)_k 
U(k+1,2-m,1),
\label{eq:Pketa0}
\end{equation}

\noindent
where $U(a,b,z)$ is Tricomi's confluent hypergeometric function
\cite{Olver2010}.
To obtain an asymptotic expression of the tail of 
the degree distribution $P_{\rm st}(k)$, for $k \gg 1$, we
use the asymptotic expansion
(Eq. 13.8.8 in Ref. \cite{Olver2010})

\begin{equation}
U(k+1,2-m,1)
\simeq
\frac{2 \sqrt{e}}{k!} (1-e^{-w})^{m-2} 
\sqrt{ 2 \beta \tanh \left( \frac{w}{2} \right) }
K_{m-1} [2 \beta (k+1) ],
\end{equation}

\noindent
where 

\begin{equation}
w=  \cosh^{-1} \left[ 1 + \frac{1}{2(k+1)} \right] 
\label{eq:w}
\end{equation}

\noindent
is a parameter that depends on the degree $k$,

\begin{equation}
\beta = \frac{ w + \sinh(w) }{2} 
\label{eq:beta}
\end{equation}

\noindent
is a parameter that depends on $w$,
and $K_{\nu}(x)$ is the modified Bessel function of the second kind
\cite{Olver2010}.
In the limit of $k \gg 1$ these parameters can be further approximated by
$w \simeq k^{-1/2}$ and $\beta \simeq k^{-1/2}$. Inserting these
expressions into Eq. (\ref{eq:Pketa0}) and expanding the modified
Bessel function and the Pochhammer symbol in the large $k$ limit,
we obtain

\begin{equation}
P_{\rm st}(k) \simeq \frac{ \sqrt{ \pi e }}{(m-1)!} k^{ \frac{2m-3}{4} }  e^{ - 2 \sqrt{k} }.
\label{eq:Pketa0tail}
\end{equation}

\noindent
Therefore, the tail of the stationary degree distribution is dominated 
by a stretched exponential term.
This result is consistent with the asymptotic results obtained in Refs. 
\cite{Moore2006,Ghoshal2013,Bauke2011}.

\begin{figure}
\centerline{
\includegraphics[width=7.0cm]{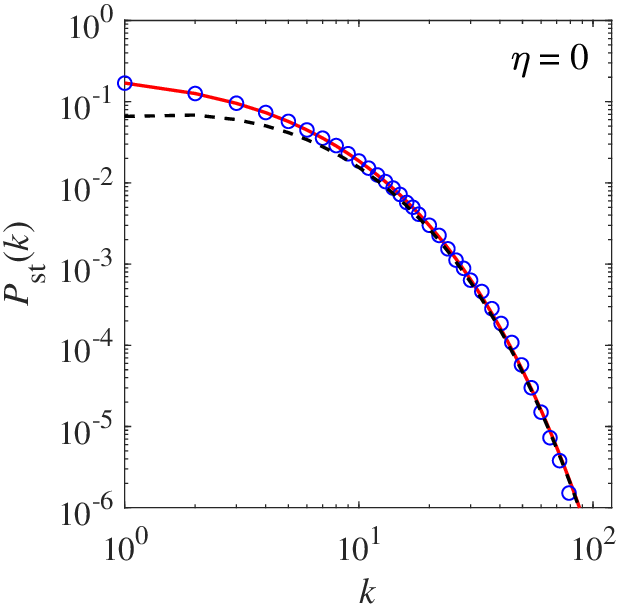} 
}
\caption{
Analytical results (solid line), 
obtained from Eq. (\ref{eq:Pketa0}),
for the stationary
degree distribution $P_{\rm st}(k)$ of a network that evolves 
via the PARD model with $m=4$
in the special case of $\eta=0$ in which the network size is fixed,
apart from possible fluctuations.
The initial network is an ER network of size $N_0=10^4$ with mean degree $c=3$.
We also present simulation results (circles)  
for the time-dependent degree distribution
$P_t(k)$, which are shown at time $t=6 N_0$.
The simulation results coincide with the analytical results,
which implies that the degree distribution has fully converged to its asymptotic form.
In the large $k$ limit, the stationary degree distribution $P_{\rm st}(k)$ converges towards
the asymptotic stretched exponential tail (dashed line), given by 
Eq. (\ref{eq:Pketa0tail}).
}
\label{fig:2}
\end{figure}

In Fig. \ref{fig:2} we present analytical
results (solid line) for the stationary
degree distribution $P_{\rm st}(k)$ of networks that evolve via the PARD dynamics 
in the special case of $\eta=0$, 
in which the expected size of the network remains fixed,
apart from stochastic fluctuations.
The initial network is an ER network of size $N_0=10^4$ with mean degree $c=3$.
The analytical results for $P_{\rm st}(k)$ 
are obtained from Eq. (\ref{eq:Pketa0}). 
The analytical results are in very good agreement with the
simulation results (circles), which are shown for $t=6 N_0$, where
the degree distribution has already converged to its asymptotic form
$P_{\rm st}(k)$.
The dashed line shows the asymptotic stretched exponential tail, given by 
Eq. (\ref{eq:Pketa0tail}).

\subsection{The case of $-1 <\eta < 0$}

The first steps in the solution of the differential equation for the
generating function $G(u)$ in the range of $-1 < \eta < 0$
are the same as shown above for $0 < \eta \le 1$.
The homogeneous part $G^{(h)}(u)$ of the solution diverges
at $u=1$ and is thus discarded.
One is left with
Eq. (\ref{eq:duGM}), which is also valid for $-1 < \eta < 0$.
Integrating Eq. (\ref{eq:duGM}) between $u$ and $1$, we obtain

\begin{equation}
G(u) M(u) = (1+\eta) \int_{u}^{1} (1-\eta-v)^{ m(1+\eta)+ \frac{1}{\eta} }
(1-v)^{ -   2 - \frac{1}{\eta} } dv,
\label{eq:GuMu}
\end{equation}

\noindent
where we relied on the fact that $M(1)=0$.
The integral on the right hand side of Eq. (\ref{eq:GuMu}) can be 
written in terms of a hypergeometric function, leading to

\begin{equation}
G(u) = \left( 1 - \frac{1-u}{\eta} \right)^{ -   m(1+\eta) - 1 - \frac{1}{\eta} }
\, _2F_1 \left[ \left.
\begin{array}{c}
-m(1+\eta) - \frac{1}{\eta}, -1 - \frac{1}{\eta} \\
- \frac{1}{\eta}
\end{array}
\right|    \frac{1-u}{\eta}
\right].
\label{eq:Gpu1eta0}
\end{equation}

\noindent
Using Eq. (\ref{eq:PtkD}) and
identity 15.5.6 from Ref. 
\cite{Olver2010},
we obtain

\begin{eqnarray}
P_{\rm st}(k) = 
&
\left( 1 - \frac{1}{\eta} \right)^{ - m(1+\eta) - 1 - \frac{1}{\eta} }
\frac{ (m+m\eta)_k }{ \left( - \frac{1}{\eta} \right)_k } 
\times
\nonumber \\
& \ \ \ \ \ 
(1-\eta)^{-k}
\, _2F_1 \left[ \left.
\begin{array}{c}
-m(1+\eta) - \frac{1}{\eta}, -1 - \frac{1}{\eta} \\
- \frac{1}{\eta} + k
\end{array}
\right|    \frac{1}{\eta}
\right].
\label{eq:Pstketaneg}
\end{eqnarray}

\noindent
In the large $k$ limit, 
the hypergeometric function on the right hand side of Eq. (\ref{eq:Pstketaneg})
is dominated by the $n=0$ term in the sum on the right hand side of Eq. (\ref{eq:2F1}),
which is equal to $1$.
Approximating the ratio between the two Pochhammer symbols on the right hand side of 
Eq. (\ref{eq:Pstketaneg}) using the Stirling formula, we obtain 

\begin{equation}
P_{\rm st}(k) \simeq 
\frac{ \Gamma \left( - \frac{1}{\eta} \right) }{\Gamma(m + m \eta)}
\left( 1 - \frac{1}{\eta} \right)^{ m(1+\eta) + 1 + \frac{1}{\eta} }
k^{ m(1+\eta) + \frac{1}{\eta} } (1-\eta)^{-k}.
\label{eq:Pstketaneg2}
\end{equation}

\noindent
Thus, the stationary degree distribution $P_{\rm st}(k)$ exhibits an exponential tail
of the form $P_{\rm st}(k) \propto e^{ - \alpha k }$,
where $\alpha = \ln (1 + |\eta|)$.

The model studied in this paper has two parameters,
namely the number $m$ of edges that are added to the network in each growth 
step and the growth/contraction rate $\eta$,
which takes values in the range of $-1 < \eta \le 1$.
It turns out that the parameter $m$ affects the density of the network 
but has little effect on the phase diagram, which is essentially determined by $\eta$.
For $\eta > 0$ the overall process is of network growth, 
while for $\eta < 0$ the overall process is of network contraction.
Thus, in the overall growth phase the life time of the network is 
unlimited, while in the overall contraction phase the network
eventually vanishes.
In the analysis presented above 
it was shown that for $\eta>0$ the
stationary degree distribution exhibits a power-law tail of the
form $P_{\rm st}(k) \sim k^{ - 2 - \frac{1}{\eta} }$.
In contrast, for $\eta < 0$ the stationary degree distribution
exhibits an exponential tail of the form 
$P_{\rm st}(k) \sim (1+|\eta|)^{-k}$.
This degree distribution remains valid as long as the
network size is not too small to support it.
The transition from a power-law tail to an exponential tail of the degree distribution,
which takes place at $\eta=0$, is a structural phase transition.
At the transition point the degree distribution exhibits a stretched exponential tail
of the form $P_{\rm st}(k) \sim \exp(-2 \sqrt{k})$,
which decays faster than a power-law but slower than an exponential.
The appearance of such an intermediate behavior is common at second order phase transitions
\cite{Schwartz2001,Schwartz2003},
thus providing further indication that there is a phase transition at $\eta=0$.

The origin of this phase transition can be traced to the structure of singularities
of Eq. (\ref{eq:diffeq0b}).
This equation exhibits two regular-singular points, one of them at $u=1$
and the other at $u=1-\eta$.
In the special case of $\eta=0$ these singular points coalesce
and the singularity at $u=1$ becomes irregular, 
giving rise to the stretched exponential tail.
The convergence of the power-series for $G(u)$
around the origin is determined by the
singularity which is closer to the origin. 
Thus, for $\eta>0$ the convergence is determined by the singularity at
$u=1-\eta$, while for $\eta<0$ it is determined by the
singularity at $u=1$.

\begin{figure}
\centerline{
\includegraphics[width=7.0cm]{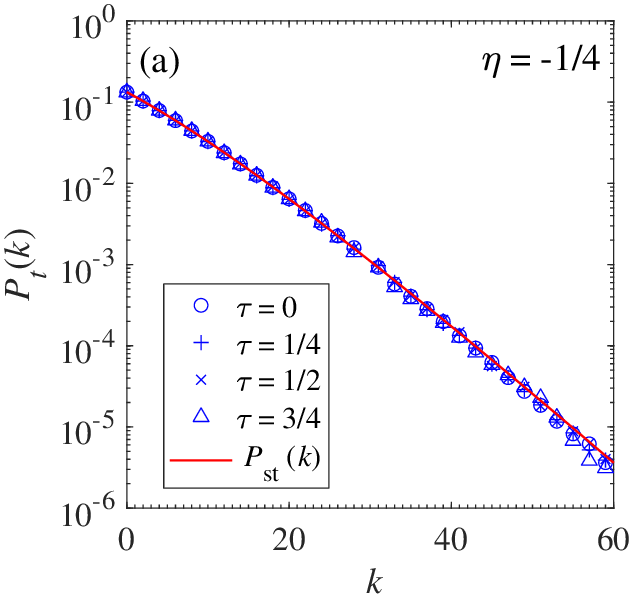}
\includegraphics[width=7.0cm]{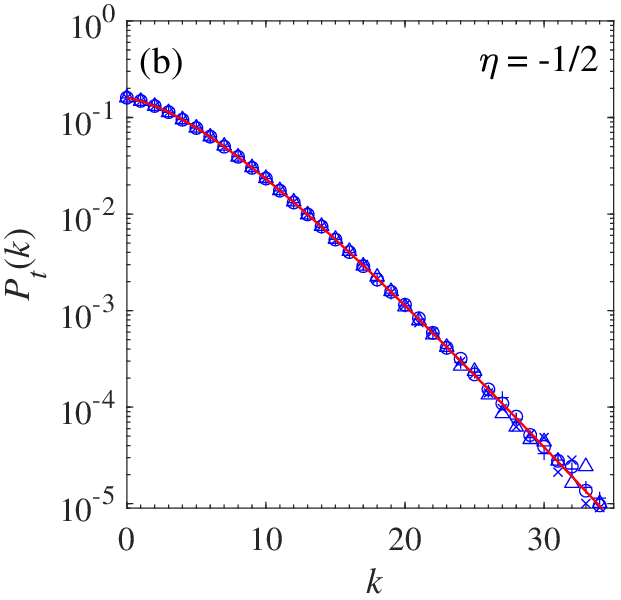}
}
\caption{
Simulation results (symbols) for the time-dependent
degree distributions $P_t(k)$
of networks that evolve  
via the PARD model with $m=8$,  
in the regime of overall network contraction
with (a) $\eta=-1/4$ and (b) $\eta=-1/2$.
The initial degree distribution is given by
$P_0(k) = P_{\rm st}(k)$, namely, it coincides with the stationary distribution
in the range of $-1 < \eta < 0$, which is given by Eq. (\ref{eq:Pstketaneg}).
The simulation results are shown for
$\tau=0$ ($\circ$),
$\tau=1/4$ ($+$),
$\tau=1/2$ ($\times$)
and $\tau=3/4$ (\scalebox{0.7}{$\triangle$}),
where the normalized time $\tau$ is the fraction 
of nodes that have been deleted [Eq. (\ref{eq:tau})].
It is found that the degree distribution remains stationary throughout the simulation
and that it coincides with the
stationary distribution $P_{\rm st}(k)$ 
(solid lines),
which is given by Eq. (\ref{eq:Pstketaneg}).
This confirms that the stationary distribution is indeed stable for
a long period of time even in the regime of overall network contraction.
}
\label{fig:3}
\end{figure}

In Fig. \ref{fig:3} we present simulation
results (symbols) for the
degree distributions of networks that evolve via the PARD dynamics
in the regime of overall network contraction
for (a) $\eta=-1/4$ and (b) $\eta=-1/2$.
The initial degree distribution is given by 
$P_0(k) = P_{\rm st}(k)$,
namely, it coincides with the stationary degree distribution
for $-1 < \eta < 0$, which is given by Eq. (\ref{eq:Pstketaneg}).
In each frame the degree distribution $P_t(k)$ 
obtained from computer simulations with these initial conditions is shown
for
$\tau=0$ ($\circ$),
$\tau=1/4$ ($+$),
$\tau=1/2$ ($\times$)
and $\tau=3/4$ (\scalebox{0.7}{$ \triangle $}),
where the normalized time $\tau$ is the fraction 
of nodes that have been deleted [Eq. (\ref{eq:tau})].
It is found that the degree distribution remains stationary throughout the simulation
and coincides with the
stationary distribution $P_{\rm st}(k)$ 
(shown by dashed lines),
which is given by Eq. (\ref{eq:Pstketaneg}).
This confirms that that the stationary distribution $P_{\rm st}(k)$ is indeed stable
even under condition in which the network contracts.
It implies that while the size of the network decreases, the local structure around a random node,
which is captured by the degree distribution, remains stationary for a finite time
window, before the network vanishes.
This result is compatible with the empirical observations reported in Ref. 
\cite{Saavedra2008}.

\begin{figure}
\centerline{
\includegraphics[width=7.0cm]{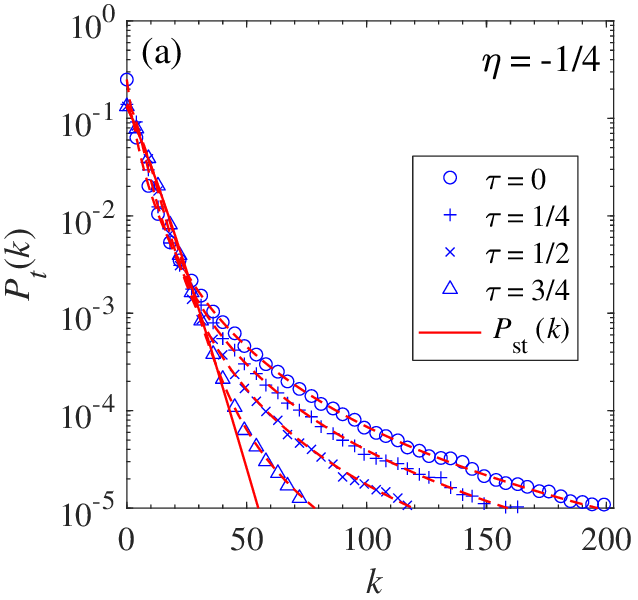}
\includegraphics[width=7.0cm]{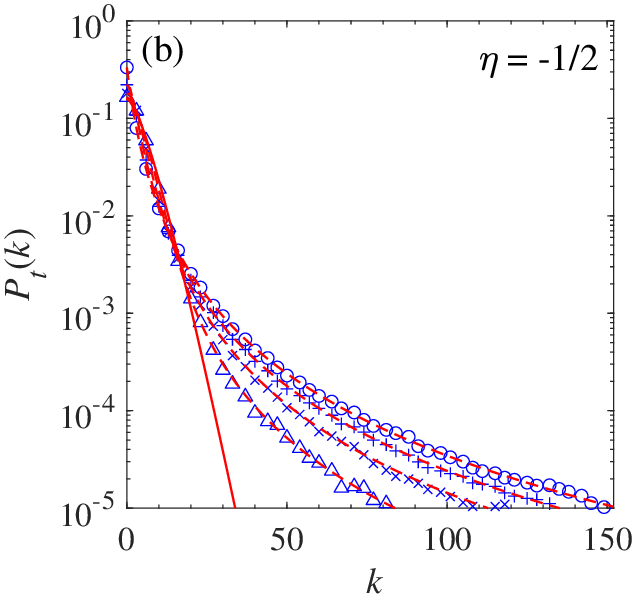}
}
\caption{
Simulation results (symbols) for the time dependent
degree distributions $P_t(k)$
of networks that evolve  
via the PARD model with $m=8$  
in the regime of overall network contraction
for (a) $\eta=-1/4$ and (b) $\eta=-1/2$.
The size of the initial networks is $N_0=20,000$ and their mean degree
satisfies $\langle K \rangle_0 = 2 \langle K \rangle_{\rm st}$,
where $\langle K \rangle_{\rm st}$ is given by Eq. (\ref{eq:Kst}).
The initial networks are generated via the PARD model with $\eta=1$.
The degree distributions $P_t(k)$ are shown
(right to left) for
$\tau=0$,
$\tau=1/4$,
$\tau=1/2$
and $\tau=3/4$,
where the normalized time $\tau$ is the fraction 
of nodes that have been deleted [Eq. (\ref{eq:tau})].
The simulation results are in very good agreement with the results
obtained from numerical integration of the master equation (dashed lines).
The stationary distributions $P_{\rm st}(k)$ are also shown
(solid lines).
As time evolves, the time dependent degree distribution $P_t(k)$ converges towards
the stationary distribution $P_{\rm st}(k)$.
The convergence becomes slower as $\eta$ is decreased.
}
\label{fig:4}
\end{figure}

In Fig. \ref{fig:4} we present
simulation results (symbols) for the
degree distributions of networks that evolve via the PARD dynamics
in the regime of overall network contraction
for (a) $\eta=-1/4$ and (b) $\eta=-1/2$.
The initial networks are generated by growth via node addition and preferential
attachment with
(a) $m_0=6$ and (b) $m_0=4$, such that the
mean degrees of the initial networks 
satisfy $\langle K \rangle_0 = 2 \langle K \rangle_{\rm st}$,
and their size is $N_0=20,000$.
In each frame the degree distribution $P_t(k)$ is shown
(right to left) for
$\tau=0$,
$\tau=1/4$,
$\tau=1/2$
and $\tau=3/4$,
where the normalized time $\tau$ is the fraction 
of nodes that have been deleted [Eq. (\ref{eq:tau})].
The simulation results are in very good agreement with the results
obtained from numerical integration of the master equation (solid lines).
The stationary distribution $P_{\rm st}(k)$ is also shown
(dashed lines).
As time evolves the time dependent degree distribution $P_t(k)$ converges towards
the stationary distribution $P_{\rm st}(k)$.
The convergence becomes slower as $\eta$ is decreased.

The mean degree $\langle K \rangle_{\rm st}=m(1+\eta)$, given by Eq. (\ref{eq:Kst}),
exhibits the same functional form for $\eta>0$, $\eta=0$ and $\eta<0$.
Below we calculate the second moment $\langle K^2 \rangle_{\rm st}$ and the
variance ${\rm Var}(K)$ of the stationary degree distribution $P_{\rm st}(k)$.
To this end, we first calculate the second factorial moment 
$\langle K(K-1) \rangle_{\rm st}$,
which is given by

\begin{equation}
\langle K(K-1) \rangle_{\rm st} = \frac{ d^2 G(u) }{d u^2} \bigg\vert_{u=1}.
\label{eq:KKm1G}
\end{equation}

\noindent
Inserting into Eq. (\ref{eq:KKm1G}) the expressions for $G(u)$ from 
Eqs. (\ref{eq:Gpu0eta1}), (\ref{eq:Gueta0}) and (\ref{eq:Gpu1eta0}) for
$\eta>0$, $\eta=0$ and $\eta<0$, respectively,
it is found that in all the three cases the second factorial moment is given by

\begin{equation}
\langle K(K-1) \rangle_{\rm st} = \frac{ 2 m (1+\eta) [ m(1+\eta) + 1 ] }{ 1 - \eta }.
\label{eq:KKm1}
\end{equation}

\noindent
Therefore, the second moment is given by

\begin{equation}
\langle K^2 \rangle_{\rm st} = \frac{ m (1+\eta) \left[ 2m(1+\eta) +3 - \eta \right] }{1 - \eta},
\end{equation}

\noindent
and the variance is given by

\begin{equation}
{\rm Var}(K) = \frac{ m (1+\eta) \left[ m (1+\eta)^2 + 3 - \eta \right] }{1 - \eta}.
\end{equation}

\noindent
The variance ${\rm Var}(K)$ is a monotonically increasing function of $\eta$.
It vanishes in the limit of $\eta \rightarrow -1^{+}$ and diverges in the limit of
$\eta \rightarrow 1^{-}$.

\section{Phase transition in the BARD model}

To examine the applicability of the results presented above beyond the specific case of the
PARD model, we calculate in Appendix B the stationary degree distribution $P_{\rm st}(k)$ of
the BARD model.
In the BARD model, at each time step, with probability $P_{\rm add}$, a BA growth step takes place in which
a new node is added
to the network and is connected preferentially to $m$ existing nodes.
Otherwise, with the complementary probability $P_{\rm del} = 1 - P_{\rm add}$
a random node is deleted together with its links.
Thus, upon formation each node connects to older nodes,
while at later times it can receive links only from younger nodes.

The stationary degree distribution of the BARD model exhibits two distinct regimes:
the regime of high degree nodes, where $k > m$ and the regime of low degree nodes, where $k < m$.
While the regime of high degree nodes consists of nodes which have gained more links than they lost
since they were added to the network, the regime of low degree nodes consists of nodes that have lost more links than they gained.
The properties of the network are mainly characterized by the degree distribution of the high degree nodes,
while the regime of low-degree nodes can be considered as a peculiarity of the model.

The degree distribution $P_{\rm st}(k)$ was calculated by Moore et al. \cite{Moore2006} in
the regime of $\eta \ge 0$.
It was found that for $\eta > 0$ it exhibits a power-law tail while for $\eta = 0$ it exhibits a stretched 
exponential tail.
Therefore, we focus in Appendix B on the case of $-1 < \eta < 0$, which has not been studied before.
In Eq. (\ref{eq:PstkBAneg}) we present an exact result for $P_{\rm st}(k)$ in this regime.
Taking the limit of $k \gg 1$ we obtain an approximation for the tail of $P_{\rm st}(k)$
and find that it exhibits an exponential decay of the form $(1-\eta)^{-k}$, 
which is identical to the exponential tail of the PARD model, given by Eq. (\ref{eq:Pstketaneg2}).
The two results differ in the algebraic dependence on $k$, which goes like $k^{\frac{1}{\eta}}$
in the BARD model, compared to $k^{m(1+\eta) + \frac{1}{\eta}}$ in the PARD model.
These results confirm the existence of a phase transition in the BARD model
between the regime of $\eta>0$ in which $P_{\rm st}(k)$ exhibits a power-law tail 
and the regime of $\eta < 0$ in which it exhibits an exponential tail.

For completeness, we also briefly consider the case of $\eta > 0$ and the special case of $\eta=0$.
The stationary degree distribution for $0 < \eta \le 1$ is given by Eq. (\ref{eq:PstkBApos}).
This result is in perfect
agreement with the corresponding results in Ref. \cite{Moore2006}.
In the special case of $\eta=0$ 
the stationary degree distribution
is given by Eqs. (\ref{eq:PstBAetaz1})
and (\ref{eq:PstBAetaz2}).
For $k \gg 1$ it can be approximated by Eq. (\ref{eq:PstBAetaz4}), and is found to exhibit a
stretched-exponential tail, in agreement with the results of Ref. 
\cite{Moore2006} and with the PARD model, where the tail of $P_{\rm st}(k)$ is given by
Eq. (\ref{eq:Pketa0tail}).

\begin{figure}
\centerline{
\includegraphics[width=7.0cm]{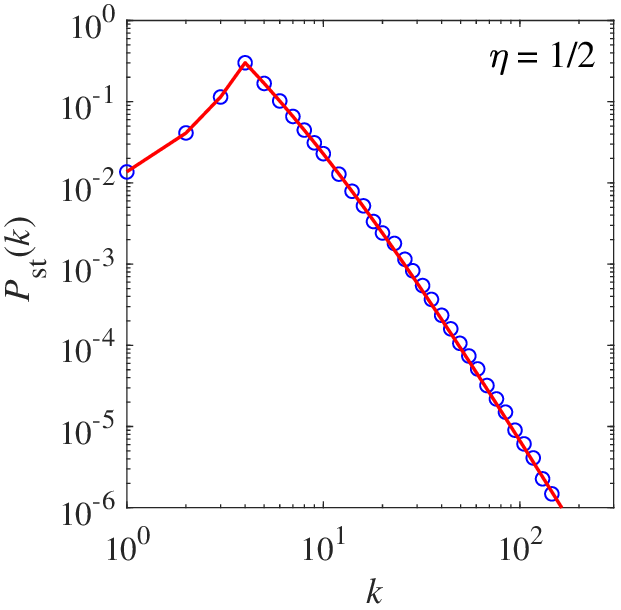} 
}
\caption{
Analytical results (solid line), 
obtained from Eq. (\ref{eq:PstkBApos}),
for the stationary
degree distributions $P_{\rm st}(k)$ of a network that evolves via the BARD model
with $m=4$,  
in the regime of overall network growth with $\eta = 1/2$.
We also present simulation results (circles) for the time-dependent degree distribution
$P_t(k)$ of a network that was grown from an initial
ER network of size $N_0=100$ with mean degree $c=3$, up to a size of $N=10^4$.
The simulation results for $P_t(k)$ coincide with the analytical results for $P_{\rm st}(k)$,
which implies that the degree distribution has fully converged to its asymptotic form.
In the large $k$ limit, the stationary degree distribution $P_{\rm st}(k)$ exhibits
a power-law tail, which is given by Eq. (\ref{eq:PstkBApos2}).
}
\label{fig:5}
\end{figure}

In Fig. \ref{fig:5} we present
analytical results (solid line), 
obtained from Eq. (\ref{eq:PstkBApos}),
for the stationary
degree distributions $P_{\rm st}(k)$ of network that evolved via the BARD model
with $m=4$,  
in the regime of overall network growth with $\eta = 1/2$.
We also present simulation results (circles) for the time-dependent degree distribution
$P_t(k)$ of a network that was grown from an initial
ER network of size $N_0=100$ with mean degree $c=3$, up to a size of $N=10^4$.
The simulation results for $P_t(k)$ coincide with the analytical results for $P_{\rm st}(k)$,
which implies that the degree distribution has fully converged to its asymptotic form.
In the large $k$ limit, the stationary degree distribution $P_{\rm st}(k)$ exhibits
a power-law tail, which is given by Eq. (\ref{eq:PstkBApos2}).
The power-law decay takes the form $k^{-2 - \frac{1}{\eta} }$, in agreement
with the PARD model, where the tail is given by Eq. (\ref{eq:Pketapostail}).

\begin{figure}
\centerline{
\includegraphics[width=7.0cm]{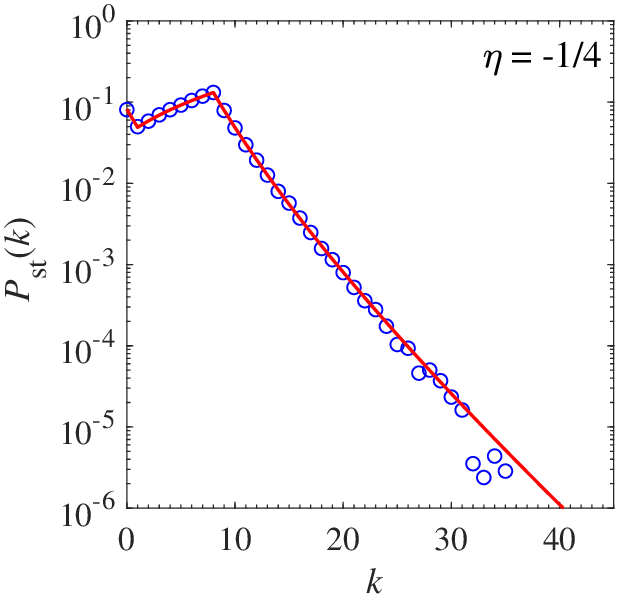} 
}
\caption{
Analytical results (solid line), 
obtained from Eq. (\ref{eq:PstkBAneg}),
for the stationary
degree distribution $P_{\rm st}(k)$ of a network that evolves via the BARD model 
with $m=8$,  
in the regime of overall network contraction with $\eta = - 1/4$.
We also present simulation results (circles) for the time-dependent degree distribution
$P_t(k)$ of a network that evolved starting from an initial
ER network of size $N_0=20,000$ with mean degree $c=8$, down to a final size of $N=2,000$.
The simulation results for $P_t(k)$ coincide with the analytical results for $P_{\rm st}(k)$,
which implies that the degree distribution has fully converged to its asymptotic form.
In the large $k$ limit, the stationary degree distribution $P_{\rm st}(k)$ exhibits an 
exponential tail, which is given by Eq. (\ref{eq:PstkBAneg2}).
}
\label{fig:6}
\end{figure}

In Fig. \ref{fig:6} we present
analytical results (solid line), 
obtained from Eq. (\ref{eq:PstkBAneg}),
for the stationary
degree distribution $P_{\rm st}(k)$ of a network that evolved via the BARD model  
with $m=8$,  
in the regime of overall network contraction with $\eta = - 1/4$.
We also present simulation results (circles) for the time-dependent degree distribution
$P_t(k)$ of a network that evolved starting from an initial
ER network of size $N_0=20,000$ with mean degree $c=8$, down to a final size of $N=2,000$.
The simulation results for $P_t(k)$ coincide with the analytical results for $P_{\rm st}(k)$,
which implies that the degree distribution has fully converged to its asymptotic form.
In the large $k$ limit, the stationary degree distribution $P_{\rm st}(k)$ exhibits an 
exponential tail, which is given by Eq. (\ref{eq:PstkBAneg2}).
The left-most data point, for $k=0$ represents the fraction of isolated nodes
in the network. These are nodes that have lost all their connections due to 
the deletion of all their neighbors.
Note that in the BARD model there no mechanism that would enable such 
isolated nodes to reconnect to other nodes.

\section{Discussion}

The process of network growth via node addition and preferential attachment
gives rise to networks that exhibit a power-law degree distribution.
Similarly, for a broad class of initial network structures
the process of network contraction via random node deletion
leads to networks that converge towards the ER structure, which
exhibits a Poisson degree distribution whose mean degree decreases
as time proceeds
\cite{Tishby2019,Tishby2020}.
The combination of growth via node addition and preferential attachment
and contraction via random node deletion yields novel structures which
depend on the balance between the rates of the two processes. 

The behavior of the degree distribution $P_t(k)$ in the regime
of overall network contraction $-1 < \eta < 0$ can be considered in the 
context of dynamical processes that exhibit intermediate asymptotic states
\cite{Barenblatt1996,Barenblatt2003}.
These are states that appear at intermediate time scales,
which are sufficiently long for such structures to build up,
but shorter than the time scales at which the whole system
disintegrates. 
During these intermediate times, the degree distribution $P_t(k)$ converges towards
the stationary degree distribution $P_{\rm st}(k)$.
The intermediate time scales can be made arbitrarily long by increasing
the initial size of the system, justifying the term `asymptotic'.

The results obtained here for the PARD and BARD models are qualitatively
different from those obtained earlier for the RARD model in which the growth process 
involves random attachment  
\cite{Budnick2022}.
In the limit of pure growth ($\eta=1$), preferential attachment gives rise to 
a power-law degree distribution with $\gamma=3$, while random attachment
gives rise to an exponential degree distribution.
Upon inclusion of the random node deletion process, in the RARD model the
exponential degree distribution is replaced by a Poisson-like distribution.
In contrast, the PARD and BARD models maintain the power-law degree distribution all the
way down to $\eta=0$. 
It implies that networks that grow by preferential attachment are more robust to 
random node deletion than networks that grow by random attachment.

The PARD and BARD models reveal intermediate-asymptotic states that persist 
within a finite time window, until the network vanishes. 
This behavior is in contrast with the 
pure deletion ($\eta=-1$) scenarios explored in Refs.  
\cite{Tishby2019,Tishby2020},
where no steady state is reached. 
Instead, under pure deletion the degree distribution $P_t(k)$ evolves 
such that the mean degree $\langle K \rangle_t$ diminishes in tandem with the network itself. 
Surprisingly, despite lacking a target steady-state, an out-of-equilibrium 
Poisson degree distribution spontaneously arises. 
The underlying principle behind this emergent dynamical 
state is that it represents a maximum entropy network, whose mean degree  
$\langle K \rangle_t$ evolves according to the specific deletion scenario.
Under such conditions, the initial structure is quickly forgotten, 
yielding a dynamically emergent state dominated by maximum entropy.

In light of the ubiquity of the preferential attachment mechanism and the
inevitability of the loss of some nodes due to failures or attacks, it is 
likely that the evolution of some real world networks is dominated by
processes that resemble the dynamical rules of the PARD model. 
In such cases, one may be able to
use information on the instantaneous structure of a network at a given moment
to gain insight into the dynamical processes that have shaped the network. 
For example, if the degree distribution of an evolving network exhibits a power-law tail,
one may infer that the network is likely to be in the overall growth phase.
In contrast, if the tail of the degree distribution exhibits an exponential decay,
one may infer that the network is likely to be in the overall contraction phase.
More specifically, in the case of a growing network, 
in which the degree distribution exhibits a power-law tail
of the form $k^{- \gamma}$, one may be able to 
estimate the balance between the rates of the node addition and node deletion processes
from the value of the exponent $\gamma$.

\section{Summary}

We presented analytical results for the 
stationary degree distribution $P_{\rm st}(k)$  
of networks that evolve via a combination of
growth, via node addition and preferential attachment,  
and contraction, via random node deletion.
To demonstrate the convergence towards the stationary state,
we also presented  
results for the time-dependent degree distribution $P_t(k)$, obtained from direct numerical integration 
of the master equation and from computer simulations.

In case that the rate of node addition exceeds the rate of node deletion ($\eta > 0$),
the overall process is of network growth, while in the opposite case ($\eta < 0$)
the overall process is of network contraction.
Using the master equation and the generating function formalism  
we obtained a closed form expression for the stationary degree distribution $P_{\rm st}(k)$,
which is reached at long times.
The convergence towards a stationary degree distribution during growth or contraction 
implies that in spite of the change in the overall size of the network, the local neighborhood
around typical nodes remains the same.
The statistical properties of these local neighborhoods depend on the balance between
the growth and contraction processes and are independent of the initial structure of the network.

In the overall growth scenario the degree distribution $P_t(k)$ converges towards
a stationary distribution $P_{\rm st}(k)$, 
which exhibits a power-law tail.
The resulting network is thus a scale-free network.
In the overall contraction scenario, 
during the contraction process 
the degree distribution $P_t(k)$ converges towards 
a stationary degree distribution $P_{\rm st}(k)$ 
which exhibits an exponential tail.
The stationary distribution persists
for a finite time window, before the network vanishes.
Thus, the intermediate state characterized by $P_{\rm st}(k)$ 
plays the role of an intermediate asymptotic state.

We thus conclude that at $\eta = 0$ there is a phase transition between the phase of
overall network growth, in which the stationary degree distribution exhibits a power-law tail
and the phase of overall network contraction, in which the stationary degree distribution 
exhibits an exponential tail.
At the transition, the stationary degree distribution exhibits a
stretched exponential tail, which is characteristic of second order phase transitions
\cite{Schwartz2001,Schwartz2003}.

This work was supported by grant no. 2020720 from the United States-Israel 
Binational Science Foundation (BSF) and by the United States
National Science Foundation (NSF).

\appendix

\section{Calculation of $d^k G(u)/du^k$ in the case of $\eta=0$}

In this Appendix we prove by induction that in the case of $\eta=0$
the $k$'th derivative of the generating function obeys

\begin{equation}
\frac{ d^k G(u) }{du^k}  =
(m)_k 
\exp \left( \frac{1}{1-u} \right)
\sum_{i=0}^{k} (-1)^i \binom{k}{i} 
\Gamma \left( -m+1-i, \frac{1}{1-u} \right)
\left( \frac{1}{1-u} \right)^{m+k+i}.
\label{eq:dkGu2}
\end{equation}

\noindent
To this end, we first replace the variable $u$ by

\begin{equation}
z = \frac{1}{1-u}.
\end{equation}

\noindent
Using the chain rule, we find that

\begin{equation}
\frac{d}{du} = z^2 \frac{d}{dz}.
\label{eq:dudz}
\end{equation}

\noindent
We will also need identity 8.8.19 from Ref. \cite{Olver2010},
which states that

\begin{equation}
\frac{d}{dz} \left[ e^z \Gamma(a,z) \right] = - (1-a) e^z \Gamma(a-1,z).
\label{eq:8819}
\end{equation}

\noindent
Inserting $k=0$ in Eq. (\ref{eq:dkGu2}), it becomes identical with Eq. (\ref{eq:Gueta0})
in the main text, which provides a closed-form expression
for $G(u)$, confirming the induction assumption.
Assuming that Eq. (\ref{eq:dkGu2}) is valid up to $k$ derivatives,
we will show below that it is also valid in the case of $k+1$ derivatives.
From Eq. (\ref{eq:dudz}) we find that

\begin{equation}
\frac{d^{k+1} G(u)}{du^{k+1}} =
z^2 \frac{d}{dz} 
\frac{d^{k} G(u)}{du^{k}}.
\label{eq:dk1Gu}
\end{equation}

\noindent
Using the induction assumption for $k$, the derivative
${d^{k} G(u)}/{du^{k}}$ in Eq. (\ref{eq:dk1Gu})
can be replaced by the right hand side of Eq. (\ref{eq:dkGu2}).
We obtain

\begin{equation}
\frac{d^{k+1} G(u)}{du^{k+1}} =
z^2  
(m)_k 
\sum_{i=0}^{k} (-1)^i \binom{k}{i} 
\frac{d}{dz}
\left[ e^z \Gamma \left( -m+1-i, z \right)
z^{m+k+i} \right].
\end{equation}

\noindent
Carrying out the derivative 
with respect to $z$
and using Eq. (\ref{eq:8819}),
we obtain

\begin{eqnarray}
& \frac{d^{k+1} G(u)}{du^{k+1}} =
\nonumber \\
& \ \ 
(m)_k \sum_{i=0}^{k} (-1)^i \binom{k}{i} 
(m+k+i) z^{m+k+1+i} e^z \Gamma(-m+1-i,z)
\nonumber \\
& \ \ 
+ 
(m)_k \sum_{i=1}^{k+1} (-1)^i \binom{k}{i-1} 
(m+i-1) z^{m+k+1+i} e^z \Gamma(-m+1-i,z).
\end{eqnarray}

\noindent
Separating the first and the last terms in the summations,
we obtain

\begin{eqnarray}
& \frac{d^{k+1} G(u)}{du^{k+1}} = 
(m)_k    \binom{k}{0} 
(m+k) z^{m+k+1} e^z \Gamma(-m+1,z)
\nonumber \\
& \ \ \ 
+ 
(m)_k \sum_{i=1}^{k} (-1)^i 
\left[ \binom{k}{i} (m+k+i) + \binom{k}{i-1} (m+i-1) \right] \times
\nonumber \\
& \ \ \ \ \ \ \ \ 
z^{m+k+1+i} e^z \Gamma(-m+1-i,z)
\nonumber \\
& \ \ \ 
+ 
(m)_k   (-1)^{k+1} \binom{k}{k} 
(m+k) z^{m+2k+2} e^z \Gamma(-m-k,z).
\label{eq:dk1GuA}
\end{eqnarray}

\noindent
The terms in the square parentheses on the right hand side of
Eq. (\ref{eq:dk1GuA}) can be written in the form

\begin{equation}
\binom{k}{i} (m+k+i) + \binom{k}{i-1} (m+i-1) =
\binom{k+1}{i} (m+k).
\end{equation}

\noindent
Using this result together with the identity

\begin{equation}
\binom{k}{k} = 1 = \binom{k+1}{k+1},
\end{equation}

\noindent
for the binomial coefficients
and the identity
$(m)_k \times (m+k) = (m)_{k+1}$
for the Pochhammer symbols,
we obtain the equation

\begin{eqnarray}
\frac{d^{k+1} G(u) }{du^{k+1}}  &=&
(m)_{k+1} \exp \left( \frac{1}{1-u} \right)
\sum_{i=0}^{k+1} (-1)^i \binom{k+1}{i} 
\times
\nonumber \\
& & \Gamma \left( -m+1-i, \frac{1}{1-u} \right)
\left( \frac{1}{1-u} \right)^{m+k+1+i},
\label{eq:dkGu3}
\end{eqnarray}

\noindent
which is the desired result.

\section{The stationary degree distribution of the BARD model}

In the BARD model, the equation for time evolution of $N_t(k)$ differs from the 
corresponding equation for the PARD model [Eq. (\ref{eq:DeltaNtkde0})]. It takes the form

\begin{eqnarray}
\frac{d N_t(k)}{dt}  &=&
P_{\rm add}    \delta_{k,m} + m  P_{\rm add} \frac{ (k-1) N_t(k-1) - k N_t(k) }{ \langle K \rangle_t N_t }  
\nonumber \\
&+& P_{\rm del}   \frac{ (k+1) N_t(k+1) - k N_t(k) }{N_t} -  P_{\rm del}  \frac{N_t(k)}{N_t}.
\label{eq:NtkBA}
\end{eqnarray}

\noindent
The first term on the right hand side of Eq. (\ref{eq:NtkBA}) accounts for the addition of 
a new node, whose initial degree is always equal to $m$.
The second term accounts for the increased degree of existing nodes that gain 
a link from the new node, in a preferential manner.
The third term accounts for the loss of links of nodes that were connected to the deleted node,
while the fourth term accounts for the loss of the deleted node.

Inserting the time derivative of $N_t(k)$ from Eq. (\ref{eq:NtkBA}) into
Eq. (\ref{eq:dPt_Nt}), we obtain the master equation

\begin{eqnarray}
\frac{d P_t(k)}{dt} &=& 
\frac{ 1 + \eta }{2 N_t} \left[ \delta_{k,m} - P_t(k) \right]
+ \frac{ m (1+\eta) }{2 \langle K \rangle_t N_t} \left[ (k-1) P_t(k-1) - k P_t(k) \right]
\nonumber \\
&+& \frac{ 1 - \eta }{2 N_t} \left[ (k+1) P_t(k+1) - k P_t(k) \right],
\label{eq:PtkBA}
\end{eqnarray}

\noindent
where the probabilities $P_{\rm add}$ and $P_{\rm del}$ are expressed
in terms of $\eta$, using Eqs. (\ref{eq:Padd}) and (\ref{eq:Pdel}).
To obtain the stationary degree distribution $P_{\rm st}(k)$ we set 
$d P_t(k)/dt = 0$.
Multiplying Eq. (\ref{eq:PtkBA}) by $u^k$ and summing up over $k$, we obtain
a differential equation for the generating function $G(u)$ of 
the stationary degree distribution $P_{\rm st}(k)$, 
which is given by

\begin{equation}
(\eta + u - 1) (1-u) \frac{d G(u)}{du} + (1+\eta) G(u) = (1+\eta) u^m.
\label{eq:GuBA}
\end{equation}

\noindent
Inspecting Eq. (\ref{eq:GuBA}) we observe that it exhibits two singular points,
at $u=1$ and $u=1-\eta$. These points are identical to the singularities of the
corresponding equation for the PARD model [Eq. (\ref{eq:diffeq0b})].
This implies that the phase diagrams of the two models share the same basic structure.
More specifically, the two singular points coincide at $\eta=0$, implying a structural 
transition at this point.
Using the integration factor

\begin{equation}
M(u) = 
\left( \frac{ 1 - \eta - u }{ 1 - u } \right)^{ 1 + \frac{1}{\eta} },
\end{equation}

\noindent
we rewrite Eq. (\ref{eq:GuBA}) in the form

\begin{equation}
\frac{d}{du} \left[ G(u) M(u) \right] =
- (1+\eta) u^m  \frac{ (1-\eta-u)^{\frac{1}{\eta}} }{ (1-u)^{   2 + \frac{1}{\eta}  } }.
\end{equation}

\noindent
In this analysis we follow the approach of Moore et al. 
\cite{Moore2006},
who considered the BARD model for $\eta \ge 0$.

In case that $-1 < \eta < 0$ the integration factor satisfies $M(1)=0$.
Using this result, we integrate Eq. (\ref{eq:GuBA}) in the range between $u$ and $1$
and obtain

\begin{equation}
G(u) = (1+\eta) \left( \frac{1-u}{1-\eta-u} \right)^{1+\frac{1}{\eta}}
\int_{u}^{1} {v}^m \frac{ (1-v)^{ -   2 - \frac{1}{\eta}   }}{ (1-\eta-v)^{ - \frac{1}{\eta} } } dv.
\label{eq:GuBAneg1}
\end{equation}

\noindent
Changing the integration variable from $v$ to $x=(1-v)/(1-u)$, we obtain

\begin{equation}
G(u) = - \left(1+ \frac{1}{\eta} \right) 
\left(1 - \frac{1-u}{ \eta } \right)^{ - 1 - \frac{1}{\eta} }
\int_{0}^{1} dx \left[ 1 - (1-u) x \right]^m
\frac{ x^{ - 2 - \frac{1}{\eta}  }}{ \left( 1 - \frac{1-u}{\eta} x \right)^{- \frac{1}{\eta} }}.
\label{eq:GuBAneg2}
\end{equation}

\noindent
Using the binomial expansion

\begin{equation}
\left[ 1 + (u-1)x \right]^m =
1 + \sum_{\ell=1}^{m} \binom{m}{\ell} (u-1)^{\ell} x^{\ell},
\label{eq:binBA}
\end{equation}

\noindent
we obtain

\begin{eqnarray}
G(u) &=& 
- \frac{  1+ \frac{1}{\eta}   }{
\left(1 - \frac{1-u}{ \eta } \right)^{  1+\frac{1}{\eta}   } }
\int_{0}^{1} dx  
\frac{ x^{ -\left( 2 + \frac{1}{\eta} \right) }}{ \left( 1 - \frac{1-u}{\eta} x \right)^{- \frac{1}{\eta} }}
\nonumber \\ 
&-& \frac{  1+ \frac{1}{\eta}   }{
\left(1 - \frac{1-u}{ \eta } \right)^{  1+\frac{1}{\eta}   } }
\sum_{\ell=1}^{m} \binom{m}{\ell} (u-1)^{\ell}
\int_{0}^{1} dx  
\frac{ x^{ \ell -  2 - \frac{1}{\eta}   }}{ \left( 1 - \frac{1-u}{\eta} x \right)^{- \frac{1}{\eta} }}.
\label{eq:GuBAneg3}
\end{eqnarray}

\noindent
Carrying out the first integral on the right hand side of Eq. (\ref{eq:GuBAneg3}), we obtain

\begin{equation}
G(u)  =  
1 - 
\frac{ 1+ \frac{1}{\eta} }
{ \left(1 - \frac{1-u}{ \eta } \right)^{  1 + \frac{1}{\eta}   } }
\sum_{\ell=1}^{m} \binom{m}{\ell} (u-1)^{\ell}
\int_{0}^{1} dx  
\frac{ x^{ \ell -  2 - \frac{1}{\eta}   }}{ \left( 1 - \frac{1-u}{\eta} x \right)^{- \frac{1}{\eta} }}.
\label{eq:GuBAneg4}
\end{equation}
 
\noindent
To evaluate the integral on the right hand side of Eq. (\ref{eq:GuBAneg4}), we use the identity

\begin{equation}
\int_{0}^{1} dx \frac{ x^{\beta} }{ ( 1 - z x )^{\gamma} }
= \frac{1}{(1-z)^{\gamma-1} } \frac{1}{z(\beta-\gamma+1)}
+ \frac{\beta}{z(\beta-\gamma+1)}
\int_{0}^{1} dx \frac{ x^{\beta-1} }{ ( 1 - z x )^{\gamma} },
\label{eq:GuBAneg5}
\end{equation}

\noindent
which is obtained by integration by parts.
Iterating this identity $\ell-1$ times, we obtain

\begin{eqnarray}
\int_{0}^{1} dx \frac{ x^{\beta} }{ ( 1 - z x )^{\gamma}  }
 =  
- \frac{1}{ (1-z)^{\gamma-1} }
\sum_{s=0}^{\ell-2}
\frac{1}{z^{s+1}} 
\frac{ \Gamma(\beta-\gamma+1-s)}{\Gamma(\beta-\gamma+2)}
\frac{ \Gamma(\beta+1) }{ \Gamma(\beta + 1 - s) }       &
\nonumber \\
\ \ \ \ \ \ \ \ \ 
+ 
\frac{1}{z^{\ell-1}} \frac{ \Gamma(\beta-\gamma+3-\ell) }{ \Gamma(\beta-\gamma+2) }
\frac{ \Gamma(\beta+1) }{ \Gamma(\beta+2-\ell) }
\int_{0}^{1} dx \frac{ x^{\beta - \ell + 1} }{ ( 1 - z x )^{\gamma} }.  &
\label{eq:GuBAneg6}
\end{eqnarray}

\noindent
Plugging $\beta=\ell-2-1/\eta$, $\gamma= - 1/\eta$ and
$z=(1-u)/\eta$
into Eq. (\ref{eq:GuBAneg6}), we obtain

\begin{eqnarray}
\int_{0}^{1}   & \frac{ x^{\ell-2 - \frac{1}{\eta} } }{ \left( 1 - \frac{1-u}{\eta} x \right)^{- \frac{1}{\eta} } } dx
= 
\nonumber \\
&-  \frac{1}{ \left( 1 - \frac{1-u}{\eta} \right)^{ -1 - \frac{1}{\eta} } }
\sum_{s=0}^{\ell - 2} \left( \frac{\eta}{1-u} \right)^{s+1}
\frac{  (\ell-2-s)! }{  ( \ell - 1)! }
\frac{ \Gamma \left(\ell - 1 - \frac{1}{\eta}  \right) }{ \Gamma \left( \ell -1 - \frac{1}{\eta} - s \right) }
\nonumber \\
&+  \left( \frac{\eta}{1-u} \right)^{\ell-1} 
\frac{1}{  (\ell - 1)! } \frac{ \Gamma \left( \ell - 1 - \frac{1}{\eta} \right) }{ \Gamma \left( - \frac{1}{\eta} \right) }
\int_{0}^{1} \frac{ x^{-1 - \frac{1}{\eta} } }{ \left( 1 - \frac{1-u}{\eta} x \right)^{- \frac{1}{\eta} } } dx.
\label{eq:GuBAneg7}
\end{eqnarray}

\noindent
Inserting the integral from Eq. (\ref{eq:GuBAneg7}) into Eq. (\ref{eq:GuBAneg4}),
we obtain

\begin{eqnarray}
G(u)  = 
1 + \left( 1 + \frac{1}{\eta} \right) 
\sum_{\ell=2}^{m} 
(-1)^{\ell}
\binom{m}{\ell}
\frac{ \Gamma \left(\ell - 1 - \frac{1}{\eta} \right) }{ (\ell-1)! }
\times   &
\label{eq:GuBAneg8}
\\
\ \ \ \ \ \ \ \ \ \ \ \   \sum_{s=0}^{\ell-2} 
 \eta^{s+1} 
\frac{ (\ell - 2 - s)! }{ \Gamma \left( \ell -1 - \frac{1}{\eta} - s \right) }
(1-u)^{\ell - s - 1} &
\nonumber \\
 -  C_{\eta,m} \left( 1 + \frac{1}{\eta} \right) (u-1) \left( 1 - \frac{1-u}{\eta} \right)^{ - \left( 1 + \frac{1}{\eta} \right) }
\, _2F_1 \left[ \left.
\begin{array}{c}
- \frac{1}{\eta}, - \frac{1}{\eta} \\
1 - \frac{1}{\eta}
\end{array}
\right|    \frac{1-u}{\eta}
\right], &
\nonumber
\end{eqnarray}

\noindent
where

\begin{equation}
C_{\eta,m} = 
- m \eta 
\, _2F_1 \left[ \left.
\begin{array}{c}
1-m, - \frac{1}{\eta} \\
2
\end{array}
\right|    \eta
\right].
\label{eq:Cetam}
\end{equation}

\noindent
The degree distribution $P_{\rm st}(k)$ can be extracted by inserting the generating function $G(u)$
from Eq. (\ref{eq:GuBAneg8}) into Eq. (\ref{eq:PtkD}). To carry out the differentiation of $G(u)$ 
we use identity 15.5.6 in Ref. \cite{Olver2010}.
We obtain

\begin{eqnarray}
P_{\rm st}(k) &=&
\delta_{k,0} 
+ \left( 1 + \frac{1}{\eta} \right) 
\sum_{\ell=2}^{m} \binom{m}{\ell}
(-1)^{k+\ell} 
\frac{ \Gamma \left( \ell - 1 - \frac{1}{\eta} \right)  }{  (\ell - 1)! }
\times
\nonumber \\
& & \ \ \ \ \ \ \ \ \ \ \   \sum_{s=0}^{\ell-2} \eta^{s+1}
\frac{  (\ell-2-s)!  }{ \Gamma \left( \ell -1 - \frac{1}{\eta} - s \right) }
\binom{\ell - s - 1}{k}
\nonumber  \\
&+& C_{\eta,m} 
\frac{   1 + \frac{1}{\eta}   }{  \left( 1 - \frac{1}{\eta} \right)^{   1 + \frac{1}{\eta}   } }
\left\{
 \frac{  k!  }{ \left( 1 - \frac{1}{\eta} \right)_k }
\, _2F_1 \left[ \left.
\begin{array}{c}
- \frac{1}{\eta}, - \frac{1}{\eta} \\
1 - \frac{1}{\eta} + k
\end{array}
\right|    \frac{1}{\eta}
\right]
\right.
\nonumber \\
&-& \left. (1 - \delta_{k,0}) 
\frac{(1-\eta) (k-1)! }{ \left( 1 - \frac{1}{\eta} \right)_{k-1} }
\, _2F_1 \left[ \left.
\begin{array}{c}
- \frac{1}{\eta}, - \frac{1}{\eta} \\
- \frac{1}{\eta} + k
\end{array}
\right|    \frac{1}{\eta}
\right]
\right\}
(1-\eta)^{ -k }.
\label{eq:PstkBAneg}
\end{eqnarray}

\noindent
The first term on the right hand side of Eq. (\ref{eq:PstkBAneg}) consists of the
Kronecker delta and contributes only in case that $k=0$. 
The second term contributes only in the regime of $0 \le k \le m-1$.
This is due to the binomial term $\binom{\ell-s-1}{k}$ that vanishes when
$\ell-s-1 < k$. Since $\ell - s - 1 \le m-1$ the binomial term must vanish for $k \ge m$.
The third term contributes to $P_{\rm st}(k)$ for all values of $k$ and the fourth term
contributes for any $k \ne 0$.

The tail of the degree distribution, where $k \gg 1$, can be approximated by

\begin{equation}
P_{\rm st}(k) \simeq C_{\eta,m} 
\frac{ 1 + \eta }{ \left( 1 - \frac{1}{\eta} \right)^{  1 + \frac{1}{\eta}   } }
\Gamma \left( 1 - \frac{1}{\eta} \right) 
k^{\frac{1}{\eta}} (1-\eta)^{-k},
\label{eq:PstkBAneg2}
\end{equation}

\noindent
where $C_{\eta,m}$ is given by Eq. (\ref{eq:Cetam}).
This tail exhibits an exponential decay which is similar to the corresponding tail of the PARD model
in the regime of overall network contraction,
given by Eq. (\ref{eq:Pstketaneg2}).
While the exponential decay is identical, the algebraic dependence on $k$ is slightly different,
namely if follows $k^{\frac{1}{\eta}}$ while in the PARD model it follows $k^{ m(1+\eta) + \frac{1}{\eta} }$.
Since the tail is dominated by the exponential decay, this difference has little effect.

Using a similar approach, we calculate the stationary degree distribution $P_{\rm st}(k)$
for $0 < \eta \le 1$.
We obtain

\begin{eqnarray}
P_{\rm st}(k) &=&
\delta_{k,0} (1-\eta)^m
\label{eq:PstkBApos}
\\
&-&  \left( 1 + \frac{1}{\eta} \right) 
\sum_{\ell=2}^{m} \binom{m}{\ell}
(-1)^{k+\ell} 
\frac{ \Gamma \left( \ell + 1 + \frac{1}{\eta} \right)  }{  (\ell - 1)! }
\times
\nonumber \\
& & \ \ \ \ \ \ \ \ \ \ \   \sum_{s=0}^{\ell-2}
\frac{ (-\eta)^{s+1} }{ (1-\eta)^{k-m+s+1} }
\frac{  (\ell-2-s)!  }{ \Gamma \left( \ell +1 + \frac{1}{\eta} - s \right) }
\binom{\ell - s - 1}{k}
\nonumber  \\
&-& D_{\eta,m} \left( 1 + \frac{1}{\eta} \right)
\frac{   ( 1 - \eta )^m  }{  \eta^{   1 + \frac{1}{\eta}   } }
\left\{
 \frac{  k!  }{ \left( 3 + \frac{1}{\eta} \right)_k }
\, _2F_1 \left[ \left.
\begin{array}{c}
2 + \frac{1}{\eta}, 2 + \frac{1}{\eta} \\
3 + \frac{1}{\eta} + k
\end{array}
\right|  1 -  \frac{1}{\eta}
\right]
\right.
\nonumber \\
&-& \left. (1 - \delta_{k,0}) 
\frac{ (k-1)! }{ (1-\eta) \left( 3 + \frac{1}{\eta} \right)_{k-1} }
\, _2F_1 \left[ \left.
\begin{array}{c}
2 + \frac{1}{\eta}, 2 + \frac{1}{\eta} \\
2 + \frac{1}{\eta} + k
\end{array}
\right|  1 - \frac{1}{\eta}
\right]
\right\},
\nonumber
\end{eqnarray}

\noindent
where

\begin{equation}
D_{\eta,m} = 
\frac{ m \eta }{ 2 \eta + 1}
\, _2F_1 \left[ \left.
\begin{array}{c}
1-m, 2 + \frac{1}{\eta} \\
2
\end{array}
\right|    \frac{ \eta }{\eta - 1}
\right].
\label{eq:Detam}
\end{equation}

\noindent
The first term on the right hand side of Eq. (\ref{eq:PstkBApos}) includes a
Kronecker delta and contributes only in case that $k=0$. 
The second term contributes only in the regime of $0 \le k \le m-1$ 
due to the binomial term $\binom{\ell-s-1}{k}$ that vanishes when
$\ell-s-1 < k$.  
The third term contributes to $P_{\rm st}(k)$ for all values of $k$ and the fourth term
contributes for any $k \ne 0$.

The tail of the degree distribution, where $k \gg 1$, can be approximated by

\begin{equation}
P_{\rm st}(k) \simeq
D_{\eta,m} 
\frac{ (1 + \eta) \left( 1 -  \eta  \right)^{  m-1   } }{      \eta^{ 1 + \frac{1}{\eta} }      }
\Gamma \left( 3 + \frac{1}{\eta} \right) 
k^{ -2 - \frac{1}{\eta} },
\label{eq:PstkBApos2}
\end{equation}

\noindent
where $D_{\eta,m}$ is given by Eq. (\ref{eq:Detam}).
This tail exhibits a power-law decay of the form $k^{-2 - \frac{1}{\eta}}$,
which is identical to the corresponding tail of the PARD model in phase
of overall network growth, given by Eq. (\ref{eq:Pketapostail}).
 
In the special case of $\eta=0$, we find that the stationary degree distribution 
can be expressed in the form

\begin{equation}
P_{\rm st}(0) = 1 + \sum_{\ell=2}^m \sum_{s=1}^{\ell-1}
(-1)^{s-1} \binom{m}{\ell} \frac{ (s-1)! }{ (\ell-1)! }
- e A_m \Gamma(0,1),
\label{eq:PstBAetaz1}
\end{equation}

\noindent
and

\begin{equation}
P_{\rm st}(k) =  \sum_{\ell=2}^m \sum_{s=1}^{\ell-1}
(-1)^{s-k-1} \binom{m}{\ell} \binom{s}{k}
\frac{ (s-1)! }{ (\ell-1)! }
+ A_m (k-1)! U(k+1,2,1),
\label{eq:PstBAetaz2}
\end{equation}

\noindent
for $k \ge 1$.
The coefficient $A_m$ is given by

\begin{equation}
A_m = \sum_{\ell=1}^m \frac{1}{ (\ell-1)! } \binom{m}{\ell} = m \, _1F_1[1-m;2;-1],
\label{eq:Am}
\end{equation}

\noindent
where 

\begin{equation}
_1F_1[a;b;z] = \sum_{n=0}^{\infty} \frac{ (a)_n z^n }{ (b)_n n! } 
\end{equation}

\noindent
is Kummer's confluent hypergeometric function
\cite{Olver2010}.

To obtain a simplified expression for the tail of $P_{\rm st}(k)$, we first note that
the first term on the right hand side of Eq. (\ref{eq:PstBAetaz2}) vanishes for $k \ge m$.
Using identity 13.8.8 in Ref. \cite{Olver2010}, it is found that for $k \gg m$

\begin{equation}
P_{\rm st}(k) \simeq  
\frac{ 2 \sqrt{e} }{k  (1-e^{ - w })^2 } \sqrt{ 2 \beta \tanh \left( \frac{ \beta }{ 2 } \right) }
K_{1} [ 2 \beta (k+1) ],
\label{eq:PstBAetaz3}
\end{equation}

\noindent
where 
$w$ and $\beta$ are given by Eqs. (\ref{eq:w}) and (\ref{eq:beta}), respectively, and
$K_{\nu}(x)$ is the modified Bessel function of the second kind
\cite{Olver2010}.
In the large $k$ limit, one can approximate $\beta$ and $w$ by $\beta \simeq k^{-1/2}$ and
$w \simeq k^{-1/2}$.
Using the asymptotic expansion of $K_{1}(x)$, we obtain

\begin{equation}
P_{\rm st}(k) \simeq  
A_m \sqrt{ \pi e } k^{- \frac{3}{4} } e^{ - 2 \sqrt{k} }.
\label{eq:PstBAetaz4}
\end{equation}

\noindent
This tail exhibits a stretched exponential decay, which is similar to the corresponding tail
of the PARD model in the special case of $\eta=0$, given by Eq. (\ref{eq:Pketa0tail}).
While the stretched exponential tail is identical in the two models, the algebraic dependence
on $k$ is different, namely in the BARD model it follows $k^{- \frac{3}{4} }$, while in the 
PARD model it follows $k^{ \frac{m}{2} - \frac{3}{4} }$.
Since the tail is dominated by the stretched exponential decay, this difference has little effect.

\noappendix


\section*{References}

\begin{thebibliography}{10}

\bibitem{Dorogovtsev2003}
Dorogovtsev S N and Mendes J F F 2003
{\it Evolution of Networks: From Biological Nets to the Internet and WWW}
(Oxford: Oxford University Press)

\bibitem{Latora2017}
Latora V, Nicosia V and Russo G 2017 
{\it Complex Networks: Principles, Methods and Applications}
(Cambridge: Cambridge University Press)

\bibitem{Havlin2010}
Havlin S and Cohen R 2010
{\it Complex Networks: Structure, Robustness and Function}
(New York: Cambridge University Press).

\bibitem{Newman2010}
Newman M E J 2018
{\it Networks: an Introduction, Second Edition} 
(Oxford: Oxford University Press)

\bibitem{Estrada2011}
Estrada E 2011 
{\it The structure of complex networks: theory and applications} 
(Oxford: Oxford University Press)

\bibitem{Dorogovtsev2022}
Dorogovtsev S N and Mendes J F F 2022
{\it The Nature of Complex Networks}
(Oxford: Oxford University Press)

\bibitem{Redner1998}
Redner S 1998 
How popular is your paper? An empirical study of the citation distribution
{\it Eur. Phys. J. B} {\bf 4}, 131 (1998).

\bibitem{Barabasi1999}
Barab\'{a}si A -L and Albert R 1999
Emergence of scaling in random networks
{\it Science} {\bf 286} 509  

\bibitem{Albert2002}
Albert R and Barab\'{a}si A -L 2002 
Statistical mechanics of complex networks
{\it Rev. Mod. Phys.} {\bf 74} 47  

\bibitem{Krapivsky2000}
Krapivsky P L, Redner S and Leyvraz F 2000
Connectivity of growing random networks
{\it Phys. Rev. Lett.} {\bf 85} 4629  

\bibitem{Dorogovtsev2000}
Dorogovtsev S N, Mendes J F F and Samukhin A N 2000
Structure of growing networks with preferential linking 
{\it Phys. Rev. Lett.} {\bf 85}  4633  

\bibitem{Bollobas2001}
Bollob\'as B 2001 
{\it Random Graphs, Second Edition}
(London: Academic Press)

\bibitem{Cohen2000}
Cohen R, Erez K, ben-Avraham D and Havlin S 2000
Resilience of the Internet to random breakdowns
{\it Phys. Rev. Lett.} {\bf 85} 4626  

\bibitem{Cohen2001}
Cohen R, Erez K, ben-Avraham D and Havlin S 2001
Breakdown of the Internet under intentional attack 
{\it Phys. Rev. Lett.} {\bf 86} 3682  

\bibitem{Albert2000}
Albert R, Jeong H and Barab\'{a}si A -L 2000
Error and attack tolerance of complex networks
{\it Nature} {\bf 406} 378 

\bibitem{Gao2015}
Gao J, Liu X, Li D and Havlin S 2015
Recent progress on the resilience of complex networks
{\it Energies} {\bf 8} 12187 

\bibitem{Yuan2015}
Yuan X, Shao S, Stanley H E and Havlin S 2015
How breadth of degree distribution influences network robustness:
Comparing localized and random attacks
{\it Phys. Rev. E} {\bf 92} 032122  

\bibitem{Shao2015}
Shao S, Huang X, Stanley H E and Havlin S 2015
Percolation of localized attack on complex networks 
{\it New J. Phys.} {\bf 17} 023049  

\bibitem{Havlin2015}
Havlin S, Stanley H E, Bashan A, Gao J and Kenett D Y 2015
Percolation of interdependent network of networks
{\it Chaos, Solitons \& Fractals} {\bf 72} 4  

\bibitem{Shekhtman2015}
Shekhtman L M, Shai S and Havlin S 2015
Resilience of networks formed of interdependent modular networks
{\it New J. Phys.} {\bf 17}  123007  

\bibitem{Shekhtman2016}
Shekhtman L M, Danziger M M and Havlin S 2016
Recent advances on failure and recovery in networks of networks
{\it Chaos, Solitons \& Fractals} {\bf 90} 28  

\bibitem{Yuan2016}
Yuan X, Dai Y, Stanley H E and Havlin S 2016
k-core percolation on complex networks: Comparing random, localized and targeted attacks
{\it Phys. Rev. E} {\bf 93} 062302  

\bibitem{Muro2016}
Di Muro M A, La Rocca C E, Stanley H E, Havlin S and Braunstein L A 2016
Recovery of interdependent networks
{\it Scientific Reports} {\bf 6} 22834  

\bibitem{Vaknin2017}
Vaknin D, Danziger M M and Havlin S 2017 
Spreading of localized attacks in spatial multiplex networks
{\it New J. Phys.} {\bf 19} 073037  

\bibitem{Braunstein2016}
Braunstein A, Dall'Asta L, Semerjian G and Zdeborov\'a L 2016
Network dismantling
{\it Proc. Natl. Acad. Sci. USA} {\bf 113} 12368  

\bibitem{Zdeborova2016}
Zdeborov\'a L, Zhang P and Zhou H -J 2016
Fast and simple decycling and dismantling of networks
{\it Scientific Reports} {\bf 6} 37954  

\bibitem{Molloy1995}
Molloy M and Reed B B 1995
A critical point for random graphs with a given degree sequence
{\it Rand. Struct. \& Algo.} {\bf 6} 161  

\bibitem{Molloy1998}
Molloy M and Reed B B 1998 
The size of the largest component of a random graph on a fixed degree sequence
{\it Combinatorics, Probability and Computing} {\bf 7} 295  

\bibitem{Tishby2019}
Tishby I, Biham O and Katzav E 2019 
Convergence towards an Erd{\H o}s-R\'enyi graph structure in network contraction processes
{\it Phys. Rev. E} {\bf 100} 032314 

\bibitem{Tishby2020}
Tishby I, Biham O and Katzav E 2020  
Analysis of the convergence of the degree distribution of contracting random networks 
towards a Poisson distribution using the relative entropy
{\it Phys. Rev. E} {\bf 101} 062308 

\bibitem{Erdos1959}
P. Erd{\H o}s  P and A. R\'{e}nyi A 1959
On random graphs I
{\it Publ. Math. Debrecen}  {\bf 6} 290  

\bibitem{Erdos1960}
Erd{\H o}s P and R\'{e}nyi A 1960
On the evolution of random graphs
{\it Publ. Math. Inst. Hungar. Acad. Sci.} {\bf 5} 17  

\bibitem{Erdos1961}
Erd{\H o}s P and R\'{e}nyi A 1961
On the evolution of random graphs II
{\it Bull. Inst. Internat. Statist.} {\bf 38} 343  

\bibitem{Coolen2017}
Coolen A C C, Annibale A and Roberts E 2017
{\it Generating Random Networks and Graphs}
(Oxford: Oxford University Press)


\bibitem{Moore2006}
Moore C, Ghoshal G and Newman M E J 2006
Exact solutions for models of evolving networks with addition and deletion of nodes
{\it Phys. Rev. E} {\bf 74} 036121  

\bibitem{Ghoshal2013}
Ghoshal G, Chi L and Barab\'asi A -L 2013
Uncovering the role of elementary
processes in network evolution
{\it Scientific Reports} {\bf 3} 2920  

\bibitem{Bauke2011}
Bauke H, Moore C, Rouquier J B and Sherrington D 2011
Topological phase transition in a network model with preferential
attachment and node removal
{\it Eur. Phys. J. B} {\bf 83} 519  


\bibitem{Torok2017}
T\"or\"ok J and Kert\'esz J 2017
Cascading collapse of online social networks 
{\it Scientific Reports} {\bf 7} 16743  

\bibitem{Lorincz2019}
L{\H o}rincz L, Koltai J, Gy{\H o}r A F and Tak\'acs K 2019
Collapse of an online social network: Burning social capital to create it?
{\it Social Networks} {\bf 57} 43  


\bibitem{Saavedra2008}
Saavedra S, Reed-Tsochas F and Uzzi B 2008
Asymmetric disassembly and robustness in declining networks
{\it Proc. Nat. Acad. Sci. USA} {\bf 105} 16466  

\bibitem{Budnick2022}
Budnick B, Biham O and Katzav E 2022
Structure of networks that evolve under a combination of growth and contraction
{\it Phys. Rev. E} {\bf 106} 044305  

\bibitem{Artime2022}
Artime O 2022 
Stochastic resetting in a networked multiparticle system with correlated transitions 
{\it J. Phys. A} {\bf 55} 484004 

\bibitem{Vankampen2007}
van Kampen N G 2007
{\it Stochastic Processes in Physics and Chemistry,
3rd Edition}
(Amsterdam: North Holland)

\bibitem{Gardiner2004}
Gardiner C 2004
{\it Handbook of Stochastic Methods: for Physics, Chemistry and the Natural Sciences, 3rd Edition}
(Berlin: Springer-Verlag)

\bibitem{Phillips2015}
Phillips C L, Nagle H T and Chakrabortty A 2015
{\it Digital Control System: Analysis and Design, Fourth Edition}
(Harlow: Pearson Education)

\bibitem{Bender1999}
Bender C M and Orszag S A 1999
{\it Advanced Mathematical Methods for Scientists and Engineers}
(New York: Springer)

\bibitem{Olver2010}
Olver F W J, Lozier D M, Boisvert R R  and  Clark C W 2010
{\it NIST Handbook of Mathematical Functions}
(Cambridge: Cambridge University Press)

\bibitem{Schwartz2001}
Schwartz M and Edwards S F 2001
The nature of the long-time decay at a second-order transition point
{\it Europhys. Lett.} {\bf 56} 499  

\bibitem{Schwartz2003}
Schwartz M 2003
The Fokker–Planck operator at a continuous phase transition,
{\it J. Phys. A} {\bf 36} 7507 

\bibitem{Barenblatt2003}
Barenblatt G I 2003
{\it Scaling}
(Cambridge: Cambridge University Press) 

\bibitem{Barenblatt1996}
Barenblatt G I 1996
{\it Scaling, self-similarity, and intermediate asymptotics}
(Cambridge: Cambridge University Press)


\end{thebibliography}
\end{document}